\documentclass[10pt]{article}
\usepackage{amssymb,geometry,graphicx,setspace, scalefnt, verbatim, amsfonts,amsthm,bm,float,booktabs,multirow,amsmath,mathtools,color,arydshln,subcaption}
\usepackage[left]{lineno}
\usepackage[round]{natbib}
\usepackage[colorlinks=true,linkcolor=blue, citecolor=blue, filecolor=magenta, urlcolor=blue, pdfborder={0 0 0}]{hyperref}
\usepackage{enumerate}
\usepackage{dsfont}
\usepackage{cancel}
\usepackage{mathrsfs}
\usepackage{authblk}
\usepackage{tikz}
\usepackage{titlesec}
\usetikzlibrary{positioning,fit,backgrounds,arrows.meta}

\usepackage[ruled,vlined,linesnumbered]{algorithm2e}

\newcommand{\dd}{\mathop{}\!\mathrm{d}}
\allowdisplaybreaks
\newcommand{\SetAlgoItemize}{\setlength{\leftmargini}{1em}}

\title{\vspace{-1em}Bayesian palaeoclimate reconstruction from zero-inflated count-compositional pollen data: A case study of Lago Grande di Monticchio in southern Italy}
\author[1]{Andr\'{e} F. B. Menezes}
\author[2]{Andrew C. Parnell}
\author[3]{Brian Huntley}
\author[1]{Keefe Murphy}

\affil[1]{\small Hamilton Institute and Department of Mathematics and Statistics, Maynooth University, Ireland}
\affil[2]{\small School of Mathematics and Statistics, University College Dublin, Ireland}
\affil[3]{\small Department of Biosciences, Durham University, Durham, United Kingdom}

\date{}

\begin{document}
\maketitle
\onehalfspacing

\begin{abstract}
\noindent Bayesian palaeoclimate reconstruction from fossil pollen counts relies on a modern pollen-climate calibration data set to infer the pollen-climate relationships used to reconstruct past climates.
While geographically large calibration data sets improve coverage of climate space and reduce unreliable extrapolation, they also introduce substantial heterogeneity, structural zeros, and complex pollen-climate relationships.
We propose a Bayesian modular framework for palaeoclimate reconstruction from count-compositional pollen data that addresses these challenges, and provides coherent uncertainty quantification.
The framework employs the zero-and-$N$-inflated multinomial logistic-normal distribution to describe the compositional pollen counts coupled with Bayesian additive regression tree priors to model the nonlinear effects and interactions among the climate covariates.
Inference is formulated through a cut posterior distribution that modularises the analysis into forward and reconstruction modules.
The forward module is fitted once using a large modern calibration data set and its posterior uncertainty is subsequently propagated to reconstruct climate variables from fossil pollen counts.
For the reconstruction module, we develop and compare three inverse posterior sampling schemes.
Simulation studies and empirical validation on the modern data set demonstrate that a combination of sampling importance resampling with a multiple imputation technique and a continuous uniform prior over the domain of the modern climate variables achieves the best predictive performance, with well-calibrated uncertainty quantification for the climate reconstruction.
In our motivating case study, we further illustrate the proposed methodology by reconstructing a three-dimensional climate vector from fossil pollen records collected at Lago Grande di Monticchio in southern Italy.
\\

\end{abstract}

\section{Introduction}

Palaeoclimate reconstruction seeks to infer past climate conditions using indirect proxy records, providing insights into climate variation beyond the instrumental record \citep{Bradley2015}.
Among the most widely used proxies are fossil pollen assemblages, whose taxonomic composition reflects changes in vegetation driven by climatic conditions \citep{Chevalier2020}.
These pollen assemblages are observed as multivariate counts of mutually exclusive pollen taxa collected from sediment cores extracted from lakes or bogs.
These counts are compositional in nature, as they are constrained to sum to the sample-specific total number of collected pollen grains.
In this work, we develop novel methodologies to reconstruct a three-dimensional climate vector from fossil pollen records collected at the upland Lago Grande di Monticchio site (henceforth `LGdM') in southern Italy using modern pollen counts collected across the Northern Hemisphere.

A fundamental assumption underlying pollen-based palaeoclimate reconstruction is the principle of \textit{uniformitarianism} \citep{Sweeney2018}, which posits that the ecological processes governing the relationship between pollen and climate today also operated in the past.
Under this assumption, a modern calibration data set containing observations of the same pollen taxa alongside contemporary climate variables is used to learn the pollen-climate relationships.
These relationships are subsequently used to reconstruct the unknown past climate variables at the fossil sites, at which only pollen counts are observed.
In this paper, we develop a new probabilistic framework for pollen-based palaeoclimate reconstruction that addresses several key features of the data from our motivating application and important limitations of existing approaches.
We now describe our application in greater detail, before outlining the challenges it poses for existing methodologies.

\subsection{Background to the data sets}\label{sec:data}
The methods we develop use two data sources, which we refer to as modern and fossil data sets.
The modern data set contains surface samples of compositional pollen counts, along with contemporary measurements of climate variables at the given locations, while the fossil data are collected from sediment cores taken from beneath lakes or bogs where pollen has accumulated over time.
Interest lies in inferring the unobserved climate associated with the fossil pollen counts, using both the compositional pollen counts and climate variables observed in the modern data.

The modern calibration data set we use is a larger version of the RS10 data set considered by \citet{Parnell2015}, which was itself expanded from the data set originally created by \citet{Allen2000}.
Presently, it consists of $n_m=7{,}832$ surface pollen samples for $j\in\{1,\ldots,28\}$ taxa collected across the Northern Hemisphere.
For each sampling location, instrumental records of three climate variables are available.
These variables reflect aspects of climate to which the pollen taxa are sensitive \citep{Huntley2012}, and they are:
(i) \textit{GDD5}, growing degree days above $5^\circ\mathrm{C}$, the annual sum of daily temperatures exceeding this threshold ($^\circ\mathrm{C}$ days);
(ii) \textit{MTCO}, the mean temperature of the coldest month ($^\circ\mathrm{C}$);
and (iii) \textit{AET/PET}, the ratio of actual to potential evapotranspiration.
Taken together, the three-dimensional climate vector characterises the length and warmth of the growing season (GDD5), winter severity (MTCO), and the moisture available to plants (AET/PET).
The climate variables were computed through weighted averaging of the observations from nearby weather stations over climatological periods of approximately 30 years \citep{Parnell2015}.

To illustrate the proposed palaeoclimate reconstruction methodology, we analyse fossil pollen records from LGdM, a lake situated in the crater of Monte Vulture in Basilicata, southern Italy.
These data have been previously studied by \citet{Parnell2016}, using the model of \citet{Parnell2015}, among others.
This fossil data set contains $n_f=924$ samples of compositional counts for the same $d=28$ pollen taxa as in the modern data, along with associated age estimates derived from annual lamination counts and tephrachronology \citep{Brauer2007}.
The records cover a period of $132$ thousand years, starting near the end of the Penultimate Glacial Period, spanning the Last Interglacial and Last Glacial periods, and ending in the Holocene, near the present day.

Table~\ref{tab:stats_modern_fossil_pollen} summarises some taxon-specific characteristics of the modern and fossil data sets, including the mean abundance, the empirical dispersion index, $\operatorname{DI}\lbrack Y_j\rbrack=\operatorname{Var}\lbrack Y_j\rbrack/\mathbb{E}\lbrack Y_j\rbrack$, the proportion of zero counts, and the empirical zero-inflated binomial index, $\operatorname{ZI}_b\lbrack Y_j\rbrack$ \citep{Kim2018}.
Further details of these indices are provided in \citet{Menezes2026}.
Overall, the pollen composition differs substantially between the two data sets.
For instance, \textit{Pinus}.D is the most abundant taxon in the modern data set, whereas \textit{Quercus}.D dominates among the fossil records, while many taxa, such as \textit{Carpinus}, \textit{Fagus}, and \textit{Quercus}.D exhibit a much higher proportion of zeros in the modern data than in the fossil data.
Both data sets exhibit substantial overdispersion, although dispersion is consistently lower in the fossil data, as indicated by $\operatorname{DI}\lbrack Y_j\rbrack$.
Similarly, zero-inflation is predominant in both data sets, with $63.21\%$ and $46.80\%$ of all observations being zero in the modern and fossil data, respectively.
Notably, all \textit{Pinus}.H counts are zero in the fossil data, suggesting the taxon was entirely absent under the site's climatic conditions and that these zeros are structural in nature, rather than being due to sampling variability.
This is supported by the modern data set, where \textit{Pinus}.H does not present positive counts in samples collected in or near southern Italy.

These differences are expected given that the modern calibration data set spans a wide range of climatic conditions across the Northern Hemisphere, whereas the fossil records are taken from a single location.
Although this mismatch increases heterogeneity in the calibration data, it is advantageous from a palaeoclimate reconstruction perspective.
A geographically broad calibration data set is desirable because palaeoclimate reconstruction methods typically cannot reliably extrapolate beyond the observed climate space, thereby reducing the risk that the reconstructed climate lies near the boundaries of the observed climate space.
The use of a large calibration data set, however, comes at the cost of greater heterogeneity, overdispersion, structural zeros, and complex pollen-climate relationships.
This motivates us to develop a new flexible Bayesian framework for pollen-based palaeoclimate reconstruction which addresses the shortfalls of some existing methodologies.

\begin{table}[!htb]
\centering
\caption{Taxon-specific descriptive statistics for the modern and fossil (Lago Grande di Monticchio, in parentheses) pollen data sets. The following empirical statistics are given: Abundance, i.e., the average count; $\operatorname{DI}\lbrack Y_j\rbrack$, the empirical dispersion index; \% zeros, the percentage of zero counts; $\operatorname{ZI}_b\lbrack Y_j\rbrack$, the zero-inflated binomial index.}
\label{tab:stats_modern_fossil_pollen}
\small
\begin{tabular}{lrrrr}
\hline
Taxa & Abundance & $\operatorname{DI}\lbrack Y_j\rbrack$  & \% zeros & $\operatorname{ZI}_b\lbrack Y_j\rbrack$ \\
  \hline
\textit{Abies} & 10.81 (20.04) & 122.88 (67.47) & 66.16 (30.30) & 0.66 (0.30) \\
  \textit{Alnus} & 69.92 (25.57) & 196.69 (53.23) & 20.51 (16.99) & 0.21 (0.17) \\
  \textit{Artemisia} & 38.35 (30.06) & 309.33 (34.94) & 37.79 (8.44) & 0.38 (0.08) \\
  \textit{Betula} & 140.03 (10.28) & 193.07 (25.15) & 25.11 (14.61) & 0.25 (0.14) \\
  \textit{Carpinus} & 2.27 (7.17) & 141.59 (16.91) & 91.06 (36.58) & 0.81 (0.36) \\
  \textit{Castanea} & 1.72 (0.16) & 309.46 (3.86) & 92.71 (91.77) & 0.75 (0.06) \\
  \textit{Cedrus} & 3.03 (0.02) & 470.18 (6.53) & 96.65 (99.24) & 0.92 (0.02) \\
  Chenopodiaceae & 39.33 (15.01) & 321.46 (30.10) & 37.90 (9.63) & 0.38 (0.10) \\
  \textit{Corylus} & 7.31 (4.93) & 125.16 (21.60) & 78.69 (37.45) & 0.79 (0.35) \\
  Cyperaceae & 51.24 (3.21) & 220.80 (5.02) & 26.16 (26.52) & 0.26 (0.20) \\
  \textit{Ephedra} & 1.15 (1.47) & 161.94 (4.10) & 94.11 (51.30) & 0.62 (0.25) \\
  Ericales & 14.90 (0.45) & 206.28 (2.71) & 71.82 (73.81) & 0.72 (0.09) \\
  \textit{Fagus} & 13.21 (17.93) & 162.83 (44.71) & 70.93 (6.82) & 0.71 (0.07) \\
  Gramineae & 103.08 (82.00) & 198.62 (37.19) & 10.27 (0.00) & 0.10 (0.00) \\
  \textit{Juniperus} & 17.51 (8.51) & 258.97 (37.82) & 59.82 (28.68) & 0.60 (0.28) \\
  \textit{Larix} & 1.60 (0.01) & 144.21 (1.50) & 93.25 (99.68) & 0.73 (0.00) \\
  \textit{Olea} & 10.28 (0.59) & 194.72 (6.93) & 84.98 (79.98) & 0.85 (0.23) \\
  \textit{Ostrya} & 4.68 (9.40) & 41.15 (26.26) & 73.94 (37.88) & 0.73 (0.37) \\
  \textit{Phillyrea} & 1.12 (0.73) & 96.80 (5.92) & 93.12 (79.55) & 0.60 (0.30) \\
  \textit{Picea} & 87.09 (0.30) & 256.58 (2.15) & 37.37 (81.82) & 0.37 (0.07) \\
  \textit{Pinus}.D & 225.28 (21.94) & 210.42 (24.49) & 6.66 (2.38) & 0.07 (0.02) \\
  \textit{Pinus}.H & 4.80 (0.00) & 402.60 (0.00) & 97.11 (100.00) & 0.96 (1.00) \\
  \textit{Pistacia} & 2.02 (0.27) & 193.75 (4.47) & 91.41 (89.83) & 0.78 (0.13) \\
  \textit{Quercus}.D & 77.80 (104.45) & 265.59 (83.01) & 40.04 (1.30) & 0.40 (0.01) \\
  \textit{Quercus}.E & 19.77 (3.35) & 403.79 (18.70) & 85.71 (71.32) & 0.86 (0.65) \\
  \textit{Salix} & 18.10 (0.98) & 219.01 (3.40) & 41.47 (60.39) & 0.41 (0.21) \\
  \textit{Tilia} & 1.54 (1.41) & 79.42 (8.51) & 85.16 (56.71) & 0.64 (0.29) \\
  \textit{Ulmus} & 9.52 (7.24) & 57.46 (8.85) & 60.20 (17.64) & 0.60 (0.17) \\   \hline
\end{tabular}
\end{table}

\subsection{Literature review}
Inference on palaeoclimate from pollen-based data presents several statistical challenges.
First, compositional pollen counts are typically characterised by an excess of zeros.
While some zeros arise due to sampling variability, many correspond to structural zeros which reflect the impossibility of observing particular taxa under certain climate conditions.
Second, climate is inherently multivariate, with complementary variables relating to attributes such as temperature and moisture jointly influencing vegetation composition.
Third, taxon-specific responses to climate are often nonlinear and potentially multimodal, and exhibit complex interaction effects.
These features complicate the task of palaeoclimate reconstruction and necessitate an efficient procedure for inferring past climate given the fossil counts and modern data.
Overall, an effective statistical framework should therefore accommodate the compositional nature of the data, account for structural zeros, flexibly model multivariate climate responses, and provide well-calibrated uncertainty quantification.

Bayesian methods provide a coherent framework for pollen-based palaeoclimate reconstruction because they naturally propagate uncertainty from the modern calibration data set to reconstruct the climate variables.
Inspired by classical reconstruction methods \citep{terBraak1989}, early Bayesian approaches modelled pollen counts using either independent Poisson distributions \citep{Toivonen2001} or a Dirichlet-multinomial (DM) distribution \citep{Vasko2000}, with taxon-specific parameters linked to a single climate covariate through unimodal Gaussian response functions.
Building on the DM model of \citet{Vasko2000}, \citet{Holmstrom2015} and \citet{Ilvonen2016} incorporated spatio-temporal dependence and age uncertainty, respectively.
Although the assumption of unimodal Gaussian response functions may be adequate for small regional calibration data sets and a carefully selected climate covariate with a clear pollen-climate signal, it is restrictive and limits the broader applicability of these methods.
In particular, it may provide an inadequate description for large scale calibration data sets where the pollen-climate relationships are considerably more complex, as in our application involving a three-dimensional climate vector.

In light of this, \citet{Haslett2006} proposed a modular Bayesian framework that exploits the hierarchical representation of the DM distribution to model taxon-specific climate responses using Gaussian process (GP) priors.
This substantially increased modelling flexibility, allowing multimodal response functions estimated from a large modern calibration data set spanning the Northern Hemisphere.
Their modular framework also incorporated a temporally smooth prior for a two-dimensional climate vector when reconstructing palaeoclimate from fossil pollen records at Glendalough, Ireland.
Building on this work, \citet{Parnell2015} adopted the model of \citet{SalterTownshend2012} for the pollen-climate relationships, incorporated age uncertainty using the approach in \citet{Haslett2008}, extended reconstruction to three climate variables, and inferred palaeoclimate volatility.
The underlying model in \citet{SalterTownshend2012} assumes an expert-informed nested hierarchical structure for the taxa and accounts for structural zeros by modelling the lowest levels using zero-inflated univariate distributions whose parameters depend on climate covariates through a Gaussian Markov random field (GMRF).
Another extension focused on flexible specification of the pollen-climate relationships was proposed by \citet{Bhattacharya2006}, who modelled taxa responses using Dirichlet process mixtures of the unimodal Gaussian curves proposed by \citet{Vasko2000}.
This framework was extended by \citet{Mukhopadhyay2013} to reconstruct two-dimensional climate variables while accommodating structural zeros through latent indicators within a hierarchical DM model.
More recently, \citet{Tipton2019} showed that increased flexibility in modelling taxon-specific climate responses leads to improved reconstructions.
As per \citet{Haslett2006}, their approach assumed a similar DM model with taxon-specific GP priors, but approximates the GPs as linear combinations of basis function expansions.
However, both their empirical analyses and publicly available software are restricted to reconstructing a single climate covariate, thereby limiting broader applicability, and they demonstrated their methodology using only holdout modern data without reference to any fossil pollen counts.

\subsection{Our contributions}
Despite these advances, existing methods do not simultaneously accommodate structural zeros using a proper count-compositional distribution and allow flexible specifications for the taxon-specific climate responses.
Methods based on unimodal Gaussian response functions \citep{Holmstrom2015,Ilvonen2016} impose restrictive parametric assumptions and do not accommodate structural zeros, whereas the approaches using GP priors in \citet{Haslett2006} and \citet{Tipton2019} provide more flexibility but likewise do not explicitly model structural zeros.
Although the Bayesian modular framework of \citet{Parnell2015} incorporates structural zeros using the model of \citet{SalterTownshend2012}, it relies on --- and is sensitive to --- the correct specification of an externally imposed hierarchical structure and a power-link specification for the structural zero probabilities which depends on the abundance parameters, thereby limiting its flexibility.
Moreover, by modelling the lowest levels of this hierarchy using univariate zero-inflated distributions, it does not model the compositional dependence among pollen taxa explicitly.
To overcome these limitations, we develop a new Bayesian modular framework that jointly accommodates structural zeros within a coherent count-compositional distribution while flexibly estimating taxon-specific response functions for multiple climate covariates.

While modular approaches have previously been considered in the Bayesian palaeoclimate literature \citep{Haslett2006,Parnell2015}, we formulate our framework within the cut posterior paradigm \citep{Plummer2015,Jacob2017,Frazier2025,Liu2025}, thereby providing a formal probabilistic interpretation of the resulting two-stage inference procedure.
The framework consists of a forward module and a reconstruction module.
The cut posterior has the effect of `cutting feedback': in essence, this prevents information from the fossil counts from influencing the model parameters which describe the pollen-climate relationships.
This implies that the forward module infers the model parameters governing the pollen-climate relationships using only the modern calibration data, while the uncertainty in inferring past climates is propagated through the reconstruction module.
In contrast, full Bayesian inference schemes are adopted in the palaeoclimate reconstruction methods of \citet{Holmstrom2015} and \citet{Ilvonen2016}, where both the pollen-climate relationships and the missing past climate covariates are inferred simultaneously.
Such non-modular approaches have a few potentially undesirable consequences.
Rather than estimating the pollen-climate relationships solely from modern calibration data, fully Bayesian approaches which allow the fossil assemblages (for which the corresponding climate is unknown) to influence the estimation of these relationships
are susceptible to model misspecification, as the inherent uncertainty in the inferred palaeoclimate can feed back into the estimation of the pollen-climate relationships.
Relatedly, inference on the model parameters may become sensitive to the prior on the unknown past climates, through the coupling of the parameters and past climates which our modular approach avoids.
Most importantly, full Bayesian inference for palaeoclimate reconstruction would require refitting the model for each new fossil site.
In fact, even the availability of new fossil pollen samples from an already reconstructed site would necessitate refitting.
Thus, our cut posterior framework provides a computationally advantageous workflow, as the model needs to be calibrated only once.
Thereafter, the posterior draws can be stored and reused indefinitely for any number of independent reconstruction tasks for which fossil assemblages for the same taxa are available.

A key methodological innovation of our framework is the adoption of the ZANIM-LN-BART model developed by \citet{Menezes2026} for inferring the pollen-climate relationships.
This model assumes that the pollen counts follow the mixture-based zero-and-$N$-inflated multinomial logistic-normal (ZANIM-LN) distribution, which extends the zero-and-$N$-inflated multinomial distribution introduced by \citet{Menezes2025} by incorporating logistic-normal random effects.
This ZANIM-LN distribution accommodates structural zeros, complex dependence among taxa, and unobserved latent heterogeneity.
Moreover, the taxon-specific parameters of the ZANIM-LN distribution depend on climate variables through independent Bayesian additive regression trees (BART) priors (see Section~\ref{sec:forward_module} for further details).
This specification enables flexible estimation of the taxon-specific climate response functions without imposing restrictive parametric assumptions, thereby capturing nonlinear and multimodal relationships, as well interaction effects for the climate variables.
Our model further allows for both the compositional probabilities and structural zero probabilities to be covariate-dependent.

To infer past climate variables given fossil pollen counts and modern data, we develop three sampling schemes that propagate uncertainty from the forward module to the reconstruction module.
We assume that the fossil pollen observations are conditionally independent given the model parameters, so temporal dependence and age uncertainty are not considered.
Our main approach combines a sampling importance resampling (SIR) algorithm with the multiple imputation technique suggested by \citet{Plummer2015} for sampling cut posterior distributions.
In general, specifying a prior for the missing past climate variables is not a trivial task \citep{Haslett2006}.
Our SIR algorithm employs a continuous uniform prior restricted to the domain of the modern climate variables, which we define by the convex hull.
This avoids the issue encountered by \citet{Parnell2016} whereby the obtained reconstructions produced values of the AET/PET variable outside its bounded $\lbrack 0, 1\rbrack$ interval.
The continuity and simplicity of this prior also yields some computational advantages over existing Bayesian palaeoclimate reconstruction frameworks, especially for $p\geq 3$ climate dimensions.
As an example, the model of \citet{SalterTownshend2012} provides computationally fast inverse posterior samples in a two-dimensional climate space, by leveraging the integrated nested Laplace approximation \citep{Rue2009} through the use of a GMRF prior.
However, this requires the construction of a discrete (typically irregular) grid which similarly envelopes the modern climate space, and it is well known that discretisation of the climate space suffers from the curse of dimensionality as the number of grid-points explodes as the dimension of the climate space grows.
Beyond the availability of straightforward sampling schemes to uniformly draw from a convex hull, another advantage of our approach is that our prior assigns mass over the entire modern climate domain, which discrete, grid-based approaches can only approximate through increasingly computationally onerous levels of granularity.
Thus, our approach tends to produce smoother reconstructions with fewer spikes, as borne out in our LGdM case study in Section~\ref{sec:app_monticchio}.

As alternative methods to benchmark our SIR algorithm against, we consider an elliptical slice sampling (ESS) algorithm \citep{Murray2010} for which the underlying multivariate normal prior is calibrated using the empirical moments of the modern climate data.
A similar approach has previously been applied to palaeoclimate reconstruction by \citet{Tipton2019}, providing a natural point of comparison with our SIR approach.
We further present a novel constrained ESS (cESS) algorithm based on a multivariate truncated normal distribution that restricts the underlying prior to the modern climate domain, as per our SIR approach, but using a recent smooth approximation technique \citep{Maatouk2026}.
Unlike SIR, ESS and cESS employ a na{\"{i}}ve cut algorithm that is known to exhibit convergence issues \citep{Plummer2015}.
Hence, a systematic comparison of these three sampling schemes for both real and synthetic data constitutes another contribution of this paper.

The rest of our paper is organised as follows.
Section~\ref{sec:methods} provides details on our modular Bayesian framework, which adopts a cut posterior approach, along with the constituent forward and reconstruction modules in Sections~\ref{sec:forward_module} and \ref{sec:reconstruction_module}, respectively.
We show the performance of our method on synthetic data in Section~\ref{sec:sim_studies}.
We then present our application in three phases.
First, in Section~\ref{sec:app_forward_fit}, we investigate the fit of our ZANIM-LN-BART model to the full modern data set and assess the information the forward module provides about the pollen-climate relationships.
Second, in Section~\ref{sec:app_validation}, we perform a validation experiment using a train/test split on the modern data set, in order to evaluate the performance of our framework in reconstructing the holdout modern climate variables.
Third, in Section~\ref{sec:app_monticchio}, we apply our framework to address our main goal of reconstructing the past climate variables associated with the case-study of LGdM described in Section~\ref{sec:data}.
We conclude in Section~\ref{sec:discussion} with a discussion, and outline some areas for future work.
Additional algorithmic details and further results for our case study are provided in the Supplementary Material and code with the implementations of our methodology is provided through the \textsf{R} package \texttt{zanicc}, available from \url{https://github.com/AndrMenezes/zanicc}.

\section{Methods}\label{sec:methods}

Recall that the methods we develop use two data sources: modern calibration data, consisting of pollen counts and climate variables, collected in our case across the Northern Hemisphere, and fossil data, consisting of past pollen counts only, collected from lake sediments, which in our case study are from the LGdM site in southern Italy.
The modern calibration data set consists of $n_m=7{,}832$ observations from a $d$-dimensional vector of compositional pollen taxa counts, $\mathbf{y}_i^{m} \in \mathbb{S}^d_{N_i}$, and the associated observed measurements of the climate variables, $\mathbf{x}_i^m \in \mathbb{R}^p$.
We stress that the pollen counts are count-compositional vectors defined on the discrete simplex space, $\smash{\mathbb{S}^d_{N_i} = \{\mathbf{y}_i \in (0, \ldots, N_i)^d; \sum_{j=1}^d y_{ij} = N_i\}}$.
In addition, a fossil data set is available with $n_f=924$ observations of the same $d=28$ taxa, $\mathbf{y}_s \in \mathbb{S}^d_{N_s}$, but the associated past climate variables, $\mathbf{x}_s \in \mathbb{R}^p$, are missing.
In the modern data, the total counts $N_i$ range from $74$ to $1{,}003$, while the total counts $N_s$ range from $38$ to $782$ in the fossil data.
We denote the collection of fossil counts and associated past climate variables by $\mathbf{y}=\{\mathbf{y}_s\}_{s=1}^{n_f}$ and $\mathbf{x}=\{\mathbf{x}_s\}_{s=1}^{n_f}$, respectively.
For notational clarity, we denote the modern pollen counts and associated observed climate variables by $\mathbf{y}^m=\{\mathbf{y}^m_i\}_{i=1}^{n_m}$ and $\mathbf{x}^m=\{\mathbf{x}^m_i\}_{i=1}^{n_m}$, respectively.

Our goal is to develop a new probabilistic framework to perform inference on the palaeoclimate variables $\mathbf{x}_s$ from the fossil pollen counts $\mathbf{y}_s$, using the modern calibration data set $\{\mathbf{y}^m, \mathbf{x}^m\}$ to learn the relationships between the pollen taxa and climate variables.
Unlike existing Bayesian methods for pollen-based palaeoclimate reconstruction, the proposed framework is designed to simultaneously address key challenges of our pollen-climate data sets, or count-compositional data more broadly: the compositional nature of the counts, the structural zeros, and the complex and typically multimodal relationships between the pollen compositions and the climate covariates.
Furthermore, our framework is consistent with the fundamental assumptions of pollen-based climate reconstruction methods \citep{Chevalier2020,Sweeney2018}.
In particular, we adopt the dynamic equilibrium assumption which considers that the taxa observed in the modern and fossil samples are systematically related to the climate in which they live.
We also adhere to the aforementioned uniformitarianism principle, which has two consequences:
the climate variables to be reconstructed are important determinants of the corresponding pollen taxa, and
the taxa in the modern data are the same as in the fossil data, with their ecological responses being unchanged over time.
Lastly, we note that the specified statistical model should have enough flexibility to represent complex taxon-specific ecological responses to the climate variables of interest.

\begin{figure}[!htb]
\centering
\begin{tikzpicture}[
roundnode/.style={circle, draw=black!40, fill=black!10, thick, minimum size=12.5mm},
roundnodewhite/.style={circle, draw=black!40, fill=white, thick, minimum size=12.5mm},
squarednode/.style={rectangle, draw=black!40, fill=black!5, thick, minimum size=12.5mm},
squarednodewhite/.style={rectangle, draw=black!40, fill=white, thick, minimum size=12.5mm},
plate/.style={rectangle, rounded corners, draw=black!50, thick, inner sep=0.625cm},
node distance=1.5cm
]

\node[roundnode] (f) at (0,0) {$\bm{\Psi}$};
\node[roundnodewhite]      (yobs) [left=of f] {$\mathbf{y}^{m}_i$};
\node[roundnodewhite]      (ynew) [right=of f] {$\mathbf{y}_s$};
\node[squarednodewhite] (xobs) [above=of yobs] {$\mathbf{x}^{m}_i$};
\node[roundnode]        (xnew) [above=of ynew] {$\mathbf{x}_s$};

\draw[-{Latex}, thick] (f.west) -- (yobs.east);
\draw[-{Latex}, thick, dashed] (f.east) -- (ynew.west);
\draw[-{Latex}, thick] (xobs.south) -- (yobs.north);
\draw[-{Latex}, thick] (xnew.south) -- (ynew.north);

\begin{scope}[on background layer]
\node[plate, fit=(xobs) (yobs), inner sep=0.625cm] (plate_left) {};
\node at (plate_left.north) [anchor=north] {$i\in\{1,\ldots,n_m\}$};
\node at (plate_left.north) [anchor=south, yshift=1mm] {Modern data};
\node[plate, fit=(xnew) (ynew), inner sep=0.625cm] (plate_right) {};
\node at (plate_right.north) [anchor=north] {$s\in\{1,\ldots,n_f\}$};
\node at (plate_right.north) [anchor=south, yshift=1mm] {Fossil data};
\node[plate, fit=(f), inner sep=0.4cm] (plate_f) {};
\end{scope}
\end{tikzpicture}
\caption{DAG representation of our modular probabilistic palaeoclimate reconstruction framework.
Circles in (grey) white represent (un)observed random variables, while squares represent fixed observed variables.
The parameter vector $\bm{\Psi}$ collects all quantities governing the relationships between the pollen compositions and the climate variables.
Dashed lines indicate the modularisation between the modern and fossil data sets.
The main inferential target is the unobserved $p$-dimensional past climate vector $\mathbf{x}_s$ associated with each fossil sample $\mathbf{y}_s$.}
\label{fig:dag_model}
\end{figure}
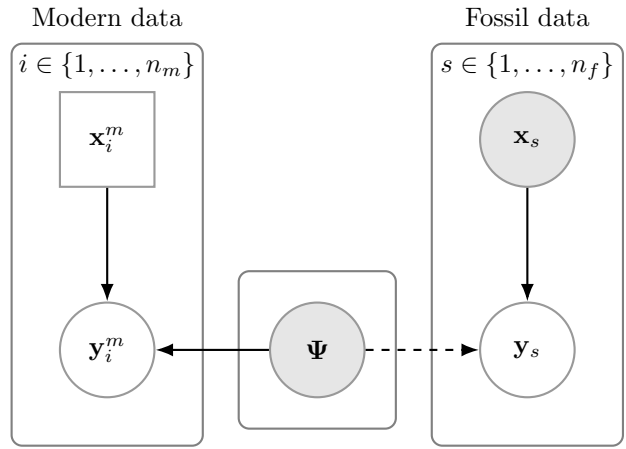

Building on these assumptions, Figure~\ref{fig:dag_model} provides a simplified representation of our framework, in the form of a directed acyclic graph (DAG).
We highlight three main features.
First, it respects the ecological causal relationship that compositional counts $\mathbf{y}_i^m$ and $\mathbf{y}_s$ are affected by their corresponding climate drivers $\mathbf{x}_i^m$ and $\mathbf{x}_s$.
Second, we assume a common set of parameters $\bm{\Psi}$ governing the relationship between pollen compositions and climate for both the modern and fossil data sets.
In Section~\ref{sec:forward_module}, we give details of the count-compositional distribution we employ, together with flexible nonparametric priors, to allow us to capture multimodal climate and structural zeros.
Finally, the dashed lines connecting the model parameters to the fossil pollen counts indicate that a modularisation occurs, as per \citet{Haslett2006} and \citet{Parnell2015}.
Note that this differs from a full probabilistic specification considered in, e.g., \citet{Holmstrom2015} and \citet{Ilvonen2016}, where the information from the fossil counts, $\mathbf{y}_s$, `feeds back' through the graph to influence inference on $\bm{\Psi}$.

From Figure~\ref{fig:dag_model}, note that the joint distribution of the random variables factorises as
\begin{equation}\label{eq:joint_likelihood}
p(\mathbf{y}, \mathbf{y}^m, \bm{\Psi}, \mathbf{x} \mid \mathbf{x}^m) =
\left\lbrack\prod_{i=1}^{n_m}p(\mathbf{y}^m_i \mid \bm{\Psi}, \mathbf{x}^m_i)
\prod_{s=1}^{n_f}p(\mathbf{y}_s \mid \bm{\Psi}, \mathbf{x}_s)\pi(\mathbf{x}_s)\right\rbrack
\times
\pi(\bm{\Psi}),
\end{equation}
where $p(\mathbf{y}^m_i \mid \bm{\Psi}, \mathbf{x}^m_i)$ and $p(\mathbf{y}_s \mid \bm{\Psi}, \mathbf{x}_s)$ are the likelihood functions of the modern and fossil compositional pollen counts, respectively, conditional on the model parameters and the climate in each case.
As the past climate variables are unobserved (indeed, they are our main inferential target) independent priors are assumed between the parameters $\bm{\Psi}$ and the past climate $\mathbf{x}_s$ i.e., $\pi(\bm{\Psi}, \mathbf{x}_s) = \pi(\bm{\Psi}) \pi(\mathbf{x}_s)$.
Further details on the priors we assume for $\pi(\bm{\Psi})$ and $\pi(\mathbf{x}_s)$ are provided in Sections~\ref{sec:forward_module} and \ref{sec:reconstruction_module}, respectively.
As per \citet{SalterTownshend2012} and \citet{Tipton2019}, we further assume that the fossil pollen counts are conditionally independent given the climate variables and parameters.
Although this assumption ignores the temporal dependence and the uncertainty in the ages of the samples, our framework can be naturally extended in future work by using a temporal prior for the age along with an age-depth chronology model
\citep{Haslett2008}.

We elaborate on the modularised nature of our approach in Section~\ref{sec:cut_posterior}, characterising the inferential target as a cut posterior distribution, before providing further details of the forward and reconstruction modules in Sections~\ref{sec:forward_module} and \ref{sec:reconstruction_module}, respectively.

\subsection{Cut posterior}\label{sec:cut_posterior}

Recall that the model parameters and unobserved past climate vectors are denoted by $\bm{\Psi}$ and $\mathbf{x}$, respectively.
Given observations from the modern data, $\{\mathbf{y}^m, \mathbf{x}^m\}$, and the fossil pollen counts, $\mathbf{y}$,
we leverage the modularisation assumption illustrated in Figure~\ref{fig:dag_model} and define the cut posterior distribution \citep{Plummer2015} as
\begin{align}
p_{\operatorname{cut}}(\mathbf{x}, \bm{\Psi} \mid \mathbf{y}^m, \mathbf{y}, \mathbf{x}^m) & \coloneqq
\underbrace{p(\bm{\Psi} \mid \mathbf{y}^m, \mathbf{x}^m)}_{\text{forward module}}
\underbrace{p(\mathbf{x} \mid \bm{\Psi}, \mathbf{y})}_{\text{reconstruction module}}  \nonumber\\
&=
\left\lbrack\frac{1}{\mathcal{Z}(\mathbf{y}^m)}\prod_{i=1}^{n_m}p(\mathbf{y}^m_i \mid \bm{\Psi}, \mathbf{x}^m_i) \pi(\bm{\Psi})
\right\rbrack \left\lbrack
\prod_{s=1}^{n_f}\frac{p(\mathbf{y}_s \mid \bm{\Psi}, \mathbf{x}_s)\pi(\mathbf{x}_s)}{\mathcal{Z}(\mathbf{y}_s \mid \bm{\Psi})}\right\rbrack,
\label{eq:cut_posterior}
\end{align}
where
$\mathcal{Z}(\mathbf{y}^m)\!=\!\int \prod_{i=1}^{n_m} p(\mathbf{y}_i^m \mid \bm{\Psi}, \mathbf{x}_i^m) \pi(\bm{\Psi}) \dd \bm{\Psi}$,
and
$\mathcal{Z}(\mathbf{y}_s \mid \bm{\Psi})\!=\!\int p(\mathbf{y}_s \mid \bm{\Psi}, \mathbf{x}_s)\pi(\mathbf{x}_s)\dd\mathbf{x}_s$
denote the normalising constants of the forward and reconstruction modules, respectively.
It is imperative to note that both likelihood functions, $p(\mathbf{y}^m_i \mid \bm{\Psi}, \mathbf{x}^m_i)$ and $p(\mathbf{y}_s \mid \bm{\Psi}, \mathbf{x}_s)$, relate to the same underlying probability model, conditional on the same parameters $\bm{\Psi}$, albeit evaluated using modern and fossil pollen counts, respectively.
We also stress that the reconstruction module comprises a $p$-dimensional distribution for the palaeoclimate variables.

Directly sampling from \eqref{eq:cut_posterior} using a Markov chain Monte Carlo (MCMC) algorithm is difficult as it requires evaluation of the term $\mathcal{Z}(\mathbf{y}_s \mid \bm{\Psi})$, which is often intractable, except for simple models.
However, the modular form of the cut posterior distribution in \eqref{eq:cut_posterior} yields simplifications which enable inference on the model parameters to be performed in two stages.
In the first stage, inference on $\bm{\Psi}$ is performed by sampling from the posterior density $p(\bm{\Psi} \mid \mathbf{y}^m, \mathbf{x}^m)$ associated to the forward module.
Conditional on the resulting posterior distribution of $\bm{\Psi}$, the second stage performs inference on $\mathbf{x}$ via the reconstruction module.
This modularisation is consistent with (though not necessitated by) the uniformitarianism principle, whereby the ecological responses of the taxa are assumed to have remained unchanged over time \citep{Sweeney2018,Chevalier2020}.
Under this assumption, the pollen-climate relationships are estimated exclusively from the modern calibration data, while the fossil counts serve only to infer past climates under the assumed invariant response functions.

Although modular inference procedures have been previously considered in Bayesian palaeoclimate reconstructions methods \citep[e.g.,][]{Haslett2006,Parnell2015,Tipton2019}, characterising them through the cut posterior approach provides a formal probabilistic interpretation of the resulting two-stage inference procedure \citep{Jacob2017}.
We note that modular inference is not limited to the field of palaeoclimate reconstruction; indeed, across many Bayesian inference problems, its theoretical properties have been well-established \citep[see, e.g.,][]{Jacob2017,Frazier2025,Liu2025,Pompe2026} and it has been adopted in a wide range of applications \citep[see, e.g.,][]{Liu2009,Styring2017}.

In our setting, the modular framework explicitly clarifies that the inference targets a valid and well-defined probability density function in which information from the fossil count data is intentionally cut in the updates of the model parameters under the forward module.
This provides a principled safeguard against a source of model misspecification that could arise if the fossil counts and corresponding reconstructed past climate covariates were allowed to influence the estimation of the pollen-climate relationships and in turn propagate to the reconstruction module.
This characterisation is also computationally important, as it highlights that the na{\"{i}}ve cut MCMC algorithm \citep{Liu2009}, which underlies some existing Bayesian palaeoclimate reconstruction methods \citep[e.g.,][]{Haslett2006,Parnell2015,Tipton2019}, does not generally converge to the cut posterior distribution in \eqref{eq:cut_posterior}, as demonstrated by \citet{Plummer2015}.
In the na{\"{i}ve cut algorithm, posterior draws $\{\bm{\Psi}^{(r)}\}_{r=1}^R$ are first obtained from the forward module, which targets $p(\bm{\Psi} \mid \mathbf{y}^m, \mathbf{x}^m)$, and each draw is subsequently used in the reconstruction module to sample $\mathbf{x}_s^{(r)} \sim p(\mathbf{x}_s \mid \mathbf{y}_s, \bm{\Psi}=\bm{\Psi}^{(r)})$.
Consequently, a single Markov chain is propagated through both modules, yielding a modified two-stage Gibbs sampler that does not generally possess the cut posterior as its stationary distribution \citep{Plummer2015} and which, without the cut feedback, does not target the full posterior distribution either \citep{Vandyk2015}.
In Section~\ref{sec:reconstruction_module}, we describe our solution to this problem based on the multiple imputation approach proposed by \citet{Plummer2015}.

\subsection{Forward module}\label{sec:forward_module}
To perform palaeoclimate reconstruction within the general modular framework outlined above, we must specify the underlying generative model for how the pollen compositions depend on the climate covariates.
In the forward module of our inferential framework, we estimate the parameters $\bm{\Psi}$ that govern these relationships using only the modern data set, $\{\mathbf{y}^m, \mathbf{x}^m\}$.
For this purpose, we adopt the ZANIM-LN-BART model of \citet{Menezes2026}, because it respects the compositional nature of the pollen counts and provides substantial flexibility to capture complex features present in our modern data set.
In particular, this model defines a proper mixture-based zero-inflated count-compositional distribution that naturally accommodates structural zeros, nonlinearities, and interactions.
Moreover, its nonparametric BART priors allow taxon-specific climate responses to be learned, with both the compositional probabilities and structural zero probabilities allowed to depend on climate covariates, without imposing restrictive parametric forms in either case.

Under the ZANIM-LN-BART model, the full parameter vector is $\bm{\Psi}=(\bm{\Psi}^{(\mathrm{c})}, \bm{\Psi}^{(0)}, \bm{\Sigma}_U)$, where $\bm{\Psi}^{(\mathrm{c})}=(\Psi^{(\mathrm{c})}_1, \ldots, \Psi^{(\mathrm{c})}_d)$ and $\bm{\Psi}^{(0)}=(\Psi^{(0)}_1, \ldots, \Psi^{(0)}_d)$ gather all taxon-specific parameters associated with the compositional probabilities and structural zero probabilities, respectively, while $\bm{\Sigma}_U$ denotes the full covariance matrix of the logistic-normal random effects.
More precisely, for each pollen taxon $j\in\{1,\ldots, d\}$, the taxon-specific parameters
$\Psi^{(\mathrm{c})}_j=\{(\mathcal{T}^{(\mathrm{c})}_{hj}, \Lambda_{hj})\}_{h=1}^{M_{\mathrm{c}}}$ and $\Psi^{(0)}_j=\{(\mathcal{T}^{(0)}_{hj}, \mathcal{M}_{hj})\}_{h=1}^{M_0}$
collect the regression tree topologies $(\mathcal{T}^{(\mathrm{c})}_{hj}, \mathcal{T}^{(0)}_{hj})$ and associated terminal node parameters $(\Lambda_{hj}, \mathcal{M}_{hj})$ for the compositional and structural zero components, respectively, where the log-linear BART of \citet{Murray2021} is employed as a prior for the former and the probit BART of \citet{Chipman2010} is employed as a prior for the latter.
The hierarchical representation of the ZANIM-LN-BART model, using only the modern data set, $\{\mathbf{y}^m, \mathbf{x}^m\}$, is then given by
\begin{align}\label{eq:multinomial}
(\mathbf{y}^m_i \mid N_i, \bm{\vartheta}_{i}, \mathbf{z}_{i}) &\sim
\begin{cases} \delta_{\mathbf{0}_d}(\cdot), & \textrm{if} \: z_{ij} = 0 \: \forall\: j\in\{1,\ldots, d\}, \\
\operatorname{Multinomial}_d\left\lbrack N_i, \vartheta_{i1}, \ldots, \vartheta_{id}\right\rbrack,
& \textrm{otherwise},
\end{cases}\\
\vartheta_{ij} &= \frac{z_{ij}f_j^{(\mathrm{c})}(\mathbf{x}^m_i; \Psi^{(\mathrm{c})}_j)e^{u_{ij}}}{\sum_{k=1}^d z_{ik}
f_k^{(\mathrm{c})}(\mathbf{x}^m_i; \Psi^{(\mathrm{c})}_k)e^{u_{ik}}}
, \label{eq:zanim_ln_bart_sr} \\
\left(z_{ij} \mid f_j^{(0)}\left(\mathbf{x}^m_i; \Psi^{(0)}_j \right) \right) &\sim \operatorname{Bernoulli}\left\lbrack
1 - \Phi\left( f_j^{(0)}\left(\mathbf{x}^m_i; \Psi^{(0)}_j \right) \right) \right\rbrack, \nonumber\\ \nonumber
\mathbf{u}_i & \sim \operatorname{Normal}_{d}\left\lbrack \mathbf{0}_{d}, \bm{\Sigma}_U\right\rbrack, \nonumber \\ \nonumber
f_j^{(\mathrm{c})} & \sim \operatorname{log-BART}\lbrack a, b, M_{\mathrm{c}}, a_\lambda\rbrack,\nonumber \\
f_j^{(0)}  &\sim \operatorname{BART}\lbrack a, b, M_0 , \sigma^{2}_{\mu} \rbrack, \nonumber
\shortintertext{where}
\log\left\lbrack f_j^{(\mathrm{c})}\left(\mathbf{x}^m_i; \Psi_j^{(\mathrm{c})}\right)\right\rbrack = \sum_{h=1}^{M_{\mathrm{c}}}g\left(\mathbf{x}^m_i; \mathcal{T}^{(\mathrm{c})}_{hj}, \Lambda_{hj}\right)\span\nonumber
\shortintertext{and}
f^{(0)}_j\left(\mathbf{x}^m_i; \Psi_j^{(0)}\right)  = \sum_{h=1}^{M_0} g\left(\mathbf{x}^m_i; \mathcal{T}^{(0)}_{hj}, \mathcal{M}_{hj}\right).\span\nonumber
\end{align}

Here, $a$ and $b$ are hyperparameters controlling the branching process prior \citep[see][for details]{Chipman2010} on both sets of tree topologies, $\mathcal{T}_{hj}^{(c)}$ and $\mathcal{T}_{hj}^{(0)}$, for the compositional and structural zero ensembles, respectively.
The corresponding numbers of trees are denoted by $M_\mathrm{c}$ and $M_0$, while $a_\lambda$ and $\sigma^{2}_{\mu}$ are hyperparameters for the priors on the corresponding terminal node parameters, $\Lambda_{hj}$ and $\mathcal{M}_{hj}$.
The covariance matrix $\bm{\Sigma}_U$ of the random effects $\mathbf{u}_i$ is not fully identifiable due to the compositional nature of the multinomial probabilities.
Identifiability is achieved through a sum-to-zero constraint on the random effects, and a nonparametric factor-analytic hyperprior is adopted for the resulting covariance structure.
All hyperparameters are specified according to the recommendations of \citet{Menezes2026}.
In particular, we set $M_\mathrm{c}=100$ and $M_0=100$.
This implies that each taxon $j\in\{1,\ldots,d\}$ has $100$ trees for each of the regression tree ensembles $f_j^{(\mathrm{c})}$ and $f_j^{(0)}$.
Bayesian inference on the model parameters is performed using the efficient MCMC algorithm developed in \citet{Menezes2026}, which leverages two steps of data augmentation to make inference amenable to adapting the Bayesian backfitting routines of \citet{Chipman2010} and \citet{Murray2021} to our setting.

We note that ZANIM-LN-BART is particularly appealing for our purposes.
It is a proper count-compositional model that captures complex pollen-climate relationships, while simultaneously accounting for structural zeros, overdispersion, and latent heterogeneity.
It was also shown by \citet{Menezes2026} to exhibit substantially better fit, in the modern data set, than competing methods which address some, but not all, of these issues.
Furthermore, it is notable that our BART-based methodology captures the complexities of the pollen-climate relationships nonparametrically and that it does so in an automatic fashion, in the sense that nonlinearities and interaction effects are detected without requiring pre-specification of the functional forms governing the dependence of the compositional and/or structural zero probabilities on the climate covariates.
This is in contrast to various existing approaches which are purely parametric and/or require strong prior information on the nature of the functional forms and/or do not account for covariate effects on the structural zero probabilities (or, moreover, do not account for structural zeros at all).

\subsection{Reconstruction module}\label{sec:reconstruction_module}

Under the cut posterior in \eqref{eq:cut_posterior}, our inferential target comprises both the parameters, $\bm{\Psi}$, and the missing past climate covariates, $\mathbf{x}$.
While inferring the parameters $\bm{\Psi}$ is the basis of the forward module presented in Section~\ref{sec:forward_module}, the reconstruction module `inverts' the estimated pollen-climate relationships in order to
infer the multivariate posterior distribution of the unobserved past climate vector $\mathbf{x}_s$ for each fossil sample $s\in\{1,\ldots,n_f\}$.
Though the reconstruction module is still governed by the same set of parameters, we stress that $\bm{\Psi}$ acts as a nuisance parameter; the main interest lies in the marginal distribution of $\mathbf{x}_s$ for $s \in \{1,\ldots,n_f\}$.
This marginal distribution is often referred to as an `inverse posterior' distribution and can be derived from the cut posterior distribution in \eqref{eq:cut_posterior} as
\begin{equation}\label{eq:marginal_cut_posterior}
p_{\operatorname{cut}}(\mathbf{x}_s \mid \mathbf{y}_s, \mathbf{y}^m, \mathbf{x}^m)
\propto\pi(\mathbf{x}_s) \int p(\mathbf{y}_s \mid \mathbf{x}_s, \bm{\Psi}) p(\bm{\Psi} \mid \mathbf{y}^m, \mathbf{x}^m) \dd \bm{\Psi}.
\end{equation}

The density in \eqref{eq:marginal_cut_posterior} is generally intractable, as the integral admits an analytical expression only for very simple models.
In fact, two main difficulties arise in our setting.
First, the posterior distribution $p(\bm{\Psi} \mid \mathbf{y}^m, \mathbf{x}^m)$ is itself only available approximately via the MCMC chain $\{\bm{\Psi}^{(r)}\}_{r=1}^R$ obtained in the forward module.
Second, the likelihood of the ZANIM-LN distribution, $p(\mathbf{y}_s \mid \mathbf{x}_s, \bm{\Psi})$, lacks a closed-form expression, owing to the presence of the logistic-normal random effects $\mathbf{u}_i$.
We overcome the latter challenge by noting that an unbiased estimate for the ZANIM-LN-BART likelihood function, denoted by $\hat{p}(\mathbf{y}_s \mid \mathbf{x}_s, \bm{\Psi})$, can be obtained from the hierarchical representation of the model in \eqref{eq:zanim_ln_bart_sr}.
Given a draw of the latent variables $z_{sj}$ and $\mathbf{u}_s$, the induced augmented likelihood $p(\mathbf{y}_s \mid N_s, \bm{\vartheta}_s, \mathbf{z}_s)$ is simply the multinomial probability mass function (PMF) in \eqref{eq:multinomial}, with $\vartheta_{sj}\propto z_{sj}f_j^{(\mathrm{c})}(\mathbf{x}_s)e^{u_{sj}}$ being the individual-level probabilities, which depend on the latent variables and both sets of taxon-specific regression tree ensembles evaluated at $\mathbf{x}_s$, and $N_s$ being the total number of pollen grains for a given fossil pollen assemblage.
Then, because the expectation of the augmented likelihood over the latent variables results in the marginal likelihood, we obtain an unbiased estimator of the ZANIM-LN-BART likelihood.
A consistent estimator can be constructed by averaging $T$ independent draws of the latent variables, via $\hat{p}(\mathbf{y}_s \mid \mathbf{x}_s, \bm{\Psi})=T^{-1} \sum_{t=1}^T p(\mathbf{y}_s \mid N_s, \bm{\vartheta}^{(t)}_s, \mathbf{z}^{(t)}_s)$, where each $p(\mathbf{y}_s \mid N_s, \bm{\vartheta}^{(t)}_s, \mathbf{z}^{(t)}_s)$ is computed using the PMF in \eqref{eq:multinomial}.

This approach to estimating the ZANIM-LN-BART likelihood forms a key component of all three of our proposed approaches for sampling from \eqref{eq:marginal_cut_posterior}.
In the sampling importance resampling algorithm proposed in Section~\ref{sec:sir}, we use $T=1$, which still preserves the unbiasedness of the estimator, because the resampling step naturally mitigates the high variance. Moreover, empirical results (not reported here, for the sake of brevity) indicate that large values of $T$ produce an inverse posterior distribution with similar shape as $T=1$ and higher effective sample size, but at substantially larger computational cost.
For the two algorithms based on elliptical slice sampling described in Section~\ref{sec:ess}, we find that $T=100$ provides a good balance between accuracy and computational efficiency.
In particular, this choice reduces the variability of the likelihood estimates and improves the mixing of the Markov chain while keeping the computational cost manageable.

We note that full details of the algorithms for all proposed approaches to sample from \eqref{eq:marginal_cut_posterior}, including all necessary subroutines, are deferred to Appendix~\ref{supp:algorithms}.

\subsubsection{Sampling importance resampling}\label{sec:sir}

Our implementation combines a sampling importance resampling (SIR) algorithm with the multiple imputation technique suggested by \citet{Plummer2015}.
Let $q(\cdot)$ be a proposal density for $\mathbf{x}_s$ for fossil sample $s$.
For each posterior draw $\bm{\Psi}^{(r)}$, $r\in\{1,\ldots,R\}$, we generate $K$ independent proposal values, $\mathbf{x}^{(k)}_s \sim q(\cdot)$, $k\in\{1,\ldots,K\}$, and compute the unnormalised importance weights
$w^{(k)}_r = \hat{p}(\mathbf{y}_s \mid \mathbf{x}^{(k)}_s, \bm{\Psi}=\bm{\Psi}^{(r)}) \pi(\mathbf{x}^{(k)}_s) / q(\mathbf{x}^{(k)}_s)$.
A single value is then resampled from $\{\mathbf{x}^{(k)}_s\}_{k=1}^K$ with probability proportional to $w^{(k)}_r$.
We recall that $\hat{p}(\mathbf{y}_s \mid \mathbf{x}^{(k)}_s, \bm{\Psi}=\bm{\Psi}^{(r)})$ is an unbiased estimate of the ZANIM-LN-BART likelihood obtained by using $T=1$ draw of the latent variables and evaluating the PMF in \eqref{eq:multinomial}.
Overall, the computational complexity is $\mathcal{O}(RK)$ per sample and $\mathcal{O}(n_fRK)$ across all fossil samples.

We adopt a multivariate continuous uniform prior over the convex hull $\mathcal{C} \subset \mathbb{R}^p$ of the modern climate vectors $\mathbf{x}^m$, so that $\pi(\mathbf{x}_s)\propto \mathds{1}_{\mathcal{C}}(\mathbf{x}_s)$.
Such an approach is feasible in our application, as the convex hull can be easily computed given that $\mathbf{x}^m$ is three-dimensional and the number of modern data observations, $n_m=7{,}832$, is not too large.
In any case, sampling uniformly over the convex hull of the modern climate requires only one pre-computation of the Delaunay triangulation, for which efficient algorithms exist \citep{Barber1996}.
Thereafter, the corresponding non-overlapping simplices are sampled from in proportion to their volume.
For unidimensional climate, we use the uniform distribution defined on the range of the corresponding modern climate.
In each case, our prior ensures that past climates lie within the support of the modern climates.

An important practical consideration of SIR is the choice of proposal distribution, $q(\cdot)$.
We take the prior as the proposal, i.e., $q(\mathbf{x}_s)=\pi(\mathbf{x}_s)$.
This simplifies the importance weights to $w^{(k)}_r = \hat{p}(\mathbf{y}_s \mid \mathbf{x}^{(k)}_s, \bm{\Psi}=\bm{\Psi}^{(r)})$.
Although using the prior as the proposal can be inefficient when the likelihood is tightly concentrated, its simplicity is attractive, and the empirical results in the simulation studies in Section~\ref{sec:sim_studies} and the validation experiment with holdout modern data in Section~\ref{sec:app_validation} show good performance in recovering the known ground truth climate vector $\mathbf{x}_s$.
We decompose our implementation of the SIR approach into two sequential tasks.
First, we generate a single set of $K$ proposal values uniformly distributed on $\mathcal{C}$, which we use across all $R$ posterior draws of $\bm{\Psi}$.
This, in turn, allows us to efficiently pre-compute and store the taxon-specific regression tree ensembles evaluated at the proposed climates across all posterior draws.
Second, these quantities are reused in the SIR algorithm for each fossil sample $s\in\{1,\ldots, n_f\}$, thereby greatly reducing the computational runtime.
A complete specification of our SIR algorithm and all necessary subroutines is given in Appendix~\ref{supp:algorithms}.

\subsubsection{Elliptical slice sampling}\label{sec:ess}

Elliptical slice sampling (ESS), introduced by \citet{Murray2010}, is an efficient, tuning-free MCMC algorithm designed to sample from targets with a multivariate Gaussian prior distribution.
In the context of Bayesian palaeoclimate reconstruction, \citet{Tipton2019} showed that ESS performs well for sampling the past climate covariates, $\mathbf{x}_s$, although their experiments considered only a single covariate and their model did not account for structural zeros.
Despite these caveats, we therefore consider the ESS algorithm as a literature benchmark, and further introduce a novel constrained ESS (cESS) approach, which restricts the Gaussian prior to the observed modern climate domain, analogous to the SIR approach described above.

Unlike the SIR algorithm, both the ESS and the cESS algorithms employ a na{\"{i}}ve cut algorithm, which recycles the single chain $\{\bm{\Psi}^{(r)}\}_{r=1}^R$ from the forward module.
This results in a computational complexity of $\mathcal{O}(R)$ for a single fossil observation and $\mathcal{O}(n_fR)$ across the entire fossil data set.
As noted by \citet{Plummer2015}, however, this strategy may introduce convergence issues because the invariant distribution changes across iterations.
The use of ESS and cESS do not remove this theoretical concern.
However, provided that the draws of $\bm{\Psi}^{(r)}$ induce moderate changes in the conditional inverse posterior, $p(\mathbf{x}_s \mid \bm{\Psi}=\bm{\Psi}^{(r)}, \mathbf{y}_s)$, both algorithms can track the evolving target efficiently in practice.

The motivation behind our cESS approach is that the Gaussian prior underlying ESS assigns positive probability to regions of the climate space that lie outside the domain of the observed modern data.
We instead consider the truncated multivariate normal (TMV) prior  $\pi(\mathbf{x}_s) \propto \exp\left\{-0.5 \left(\mathbf{x}_s - \bm{\mu}\right)^\top\bm{\Sigma}^{-1}\left(\mathbf{x}_s - \bm{\mu}\right)\right\}\mathds{1}_{\mathcal{C}}(\mathbf{x}_s)$, where the prior mean and covariance, $\bm{\mu}$ and $\bm{\Sigma}$, respectively, are set to their corresponding empirical values computed from the modern data, and $\mathcal{C}$ is a set that defines the observed climate domain.
In the case of $p=1$ climate covariate, $\mathcal{C}$ is an interval of the form $\mathcal{C}=\{x_s \in \mathbb{R}; l \leq x_s \leq u\}$, with $l$ and $u$ being the minimum and maximum of the modern climate, respectively.
On the other hand, for $p\geq2$ we consider the convex hull of the modern climate, which can be expressed in the half-space representation as
$\mathcal{C} = \{ \mathbf{x}_s \in \mathbb{R}^p: \mathbf{A} \mathbf{x}_s \leq \mathbf{b}\}$, where $\mathbf{A}$ is a $q\times p$ matrix of coefficients, with rows $\mathbf{a}_k\in\mathbb{R}^{p \times 1}$, and $\mathbf{b}$ is a $q$-dimensional vector of offsets.

The steps of the standard ESS algorithm cannot be applied directly with the TMV prior, due to the indicator function $\mathds{1}_{\mathcal{C}}(\mathbf{x}_s)$.
We use a smooth logistic approximation of the indicator function as per \citet{Souris2019}, \citet{Ray2020}, and \citet{Maatouk2026}.
Specifically, when $p=1$, we have $\mathds{1}_{\mathcal{C}}(x_s)\approx
(1 + e^{-\eta (x_s - l)})^{-1}(1 + e^{-\eta (u-x_s)})^{-1}$, while for $p\geq 2$, we have $\mathds{1}_{\mathcal{C}}(\mathbf{x}_s)\approx \prod_{k=1}^q( 1 + e^{-\eta (\mathbf{a}^\top_k \mathbf{x}_s - b_k)})^{-1}$, where $\eta>0$ controls the sharpness of the approximation.
As per \citet{Ray2020}, we fixed $\eta=50$ in our experiments, as it gives an accurate approximation of the indicator function.
However, it is important to stress that this remains an approximation: unlike SIR, the cESS approach is still liable to produce posterior draws outside the observed climate domain defined by $\mathcal{C}$.
Under the cESS approach, the conditional inverse posterior density given the draw $\bm{\Psi}^{(r)}$ is therefore
\begin{align*}
p(\mathbf{x}_s \mid \bm{\Psi}=\bm{\Psi}^{(r)}, \mathbf{y}_s) &\propto \hat{p}(\mathbf{y}_s \mid \mathbf{x}_s, \bm{\Psi}=\bm{\Psi}^{(r)}) \prod_{k=1}^q \left( 1 + e^{-\eta (\mathbf{a}^\top_k \mathbf{x}_s - b_k)}\right)^{-1} \\
&\phantom{\propto}~
\times \exp\left\{-0.5 \left(\mathbf{x}_s - \bm{\mu}\right)^\top\bm{\Sigma}^{-1}\left(\mathbf{x}_s - \bm{\mu}\right)\right\}.
\end{align*}
We recall that an unbiased estimate of the likelihood $\hat{p}(\mathbf{y}_s \mid \mathbf{x}_s, \bm{\Psi}=\bm{\Psi}^{(r)})$ is obtained by evaluating the multinomial PMF in \eqref{eq:multinomial} conditional on draws of the latent variables, $\mathbf{z}_s$ and $\mathbf{u}_s$, and both sets of regression tree ensembles evaluated at a given covariate vector $\mathbf{x}_s$.
In particular, as discussed in Section~\ref{sec:reconstruction_module}, we compute a consistent estimate for the ZANIM-LN-BART likelihood, for each draw $r$, by averaging over $T=100$ draws of the latent variables using the corresponding posterior estimates of their parameters.
The full specification of our cESS algorithm is provided in Appendix~\ref{supp:algorithms}.The ESS algorithm can be seen as a simpler special case, using a multivariate normal rather than TMV prior, without requiring evaluation of terms associated with the aforementioned logistic approximation of the indicator function.

\section{Simulation studies}\label{sec:sim_studies}

We conduct simulation studies to assess the performance and validate our modular Bayesian palaeoclimate reconstruction methods developed in Section~\ref{sec:methods}.
We vary the number of climate covariates $p\in\{2,3\}$ and
design scenarios which closely resemble our real application, which is described and conducted in Sections~\ref{sec:data} and \ref{sec:application}, respectively.
Since there is no publicly available software for Bayesian palaeoclimate reconstruction when $p>1$, we restrict attention to a comparison between the different sampling schemes (SIR, ESS, and cESS) for the reconstruction module developed in Section~\ref{sec:reconstruction_module}.
In Appendix~\ref{supp:add_sim}, we compare our methods against two existing Bayesian approaches with available software in a scenario containing a single climate covariate.

In these simulations, we use the same number of taxa as in the case study, $d=28$, and generate the climate covariates via $\mathbf{x}^m_i \sim \operatorname{Uniform}\lbrack \mathcal{C}\rbrack$, where $\mathcal{C}$ denotes the convex hull of the modern climate from our application.
The algorithm for doing so is provided in Appendix~\ref{supp:algorithms}.For the $p=2$ setting, we construct $\mathcal{C}$ using the GDD5 and MTCO variables only, while for $p=3$ we also include the AET/PET variable.
For each setting, we simulate $n_m=1{,}000$ modern observations, with the compositional counts generated from the ZANIM-LN distribution.
Generating the data via the same underlying distributional assumptions as our model allows us to control for the effects of model misspecification and focus purely on a comparison of the sampling schemes.
In particular, the counts are generated via $\mathbf{y}^m_i\sim \operatorname{Multinomial}\lbrack N_i, \bm{\vartheta}_i\rbrack$, where the total count is sampled from $N_i \sim \operatorname{Uniform}\lbrack 1000,2000\rbrack$, and the multinomial probabilities are given by
$\vartheta_{ij}=z_{ij}e^{u_{ij}}\alpha_{ij}/\sum_{k=1}^dz_{ik}e^{u_{ik}}\alpha_{ik}$, with $z_{ij}\sim \operatorname{Bernoulli}\lbrack 1- \zeta_{ij}\rbrack$, and $\mathbf{u}_{i}\sim\operatorname{Normal}_d\lbrack \mathbf{0}_d, \bm{\Sigma}_U\rbrack$, for observation $i\in\{1,\ldots, n_m\}$ and taxon $j\in\{1,\ldots, d\}$.
The covariance matrix of the random effects is specified using a factor-analytic decomposition, $\bm{\Sigma}_U = \bm{\Gamma}\bm{\Gamma}^\top + \bm{\Psi}$, with $q=21$ factors.
The entries of the $d\times q$ factor loading matrix $\bm{\Gamma}$ are simulated from independent standard uniform distributions, while the entries of the $q$-dimensional diagonal matrix $\bm{\Psi}$ are set to equally-spaced diagonal values from $0.30$ to $0.40$.
Here, $\alpha_{ij}$ and $\zeta_{ij}$ are covariate-dependent parameters that describe the abundance and structural zero probabilities, respectively.
To induce climate spaces in two and three dimensions, we use a Gaussian process functional form given by
$\log \alpha_{ij} = \beta_j + \xi_{ij}$, where $\bm{\xi}_{j}=(\xi_{1j}, \ldots, \xi_{n_mj})\sim\operatorname{Normal}_{n_m}\lbrack \mathbf{0}_{n_m}, C(\mathbf{x}^m)\rbrack$ with $C(\mathbf{x}^m)$ denoting the correlation matrix with entries $c(\mathbf{x}_i^m, \mathbf{x}_\ell^m)$.
Here, we use the exponential covariance function $c(\mathbf{x}^m_i,\mathbf{x}^m_\ell)=e^{-\delta_{i\ell}/\rho}$, where
$\delta_{i\ell}$ denotes the Euclidean distance between $\mathbf{x}^m_i$ and $\mathbf{x}^m_\ell$, and we set the length scale parameter to $\rho=5.0$.
This specification does not assume any additive regression tree specification which is the underlying assumption of our ZANIM-LN-BART model.
Finally, for the structural zero probabilities, we follow \citet{SalterTownshend2012} and define them through a power link function as $\zeta_{ij} = 1 - (\alpha_{ij}/\sum_{k=1}^d\alpha_{ik})^{\rho_j}$, here with $\rho_j\sim \operatorname{Uniform}\left\lbrack 0.01, 0.2\right\rbrack$.
The $n_f=100$ samples of the fossil data set $\{\mathbf{y}_s, \mathbf{x}_s\}_{s=1}^{n_f}$ were generated using the same recipe.

To evaluate the performance of the methods, we compute different predictive measures for a given fossil data set $\{\mathbf{y}_s, \mathbf{x}_s\}_{s=1}^{n_{f}}$ of size $n_{f}=100$, which is therefore not used to learn the parameters of the ZANIM-LN-BART model within the forward module.
We consider the mean squared prediction error:
$\operatorname{MSPE}=n^{-1}_{f}\sum_{s=1}^{n_{f}} \sum_{k=1}^p (x_{sk} - \mu_{sk})^2$,
the mean absolute error:
$\operatorname{MAE}=n^{-1}_{f}\sum_{s=1}^{n_{f}} \sum_{k=1}^p \lvert x_{ik} - m_{sk} \rvert $,
and the empirical coverage of the $95\%$ highest posterior density region (HPD):
$95\%\operatorname{coverage}=n_f^{-1}\sum_{s=1}^{n_{f}}\mathds{1}\left( \mathbf{x}_s \in \mathcal{R}_s(\alpha)\right)$, with $\alpha=0.05$, where $\mu_{sk}$, $m_{sk}$, and $\mathcal{R}_s(\alpha)$ are the posterior mean, posterior median, and HPD region for the $s$-th fossil sample, respectively.
The MSPE and MAE are widely-used metrics, easy to understand, and reward the method which has, on average, the closest posterior mean (MSPE) or median (MAE) to the unobserved climates.
An appealing characteristic of Bayesian palaeoclimate reconstruction methods is their ability to quantify uncertainty, so it is crucial to assess how well-calibrated the methods are in terms of coverage.
Here, we consider the joint $95\%$ coverage of the $p$-dimensional past climate vector, $\mathbf{x}_s$.
To compute the HPD region $\mathcal{R}_s(\alpha)$ for all $s \in \{1,\ldots,n_f\}$ fossil samples, we use the density quantile approach of \citet{Hyndman1996}.
In addition to these standard metrics, we also use a proper scoring rule which both quantifies the quality of the predictions and rewards proper estimation of the uncertainty therein.
In particular, we average the energy score (ES) over the fossil samples \citep{Gneiting2007}.
Given the posterior draws $\{\mathbf{x}_s^{(r)}\}_{r=1}^{R}$, the ES is computed for the $s$-th observation as follows:
$\operatorname{ES}(\mathbf{x}_s)=n_f^{-1}\sum_{r=1}^R \lVert \mathbf{x}^{(r)}_s - \mathbf{x}_s \rVert_2 - 1/(2n_f^2)\sum_{r=1}^R\sum_{j=1}^R \lVert
\mathbf{x}^{(r)}_s - \mathbf{x}^{(j)}_s\rVert_2$, where $\lVert \cdot \rVert_2$ denotes the Euclidean norm on $\mathbb{R}^p$.
Moreover, for each scenario, we generate six replicates and average these predictive metrics over them.

To conduct inference on the parameters $\bm{\Psi}$, estimated under the forward module of Section~\ref{sec:forward_module}, we run the MCMC chain using $10{,}000$ iterations, with the first $5{,}000$ iterations discarded as burn-in.
The resulting chain $\{\bm{\Psi}^{(r)}\}_{r=1}^R$ of size $R=5{,}000$ was then used in the reconstruction module of Section~\ref{sec:reconstruction_module} to approximate the inverse posterior distribution of each fossil sample $\mathbf{x}_s, s \in \{1,\ldots,n_f\}$.
In particular, for the SIR algorithm of Section~\ref{sec:sir}, we generate $K=2{,}000$ proposal values in order to ensure the modern climate domain is adequately represented.
All hyperparameters of the forward and reconstruction modules are specified as described in Sections~\ref{sec:forward_module} and \ref{sec:reconstruction_module}, respectively.

The results of the simulation experiments are reported in Table~\ref{tab:results_scenario2}.
The findings show that the SIR approach consistently outperforms both the ESS and cESS approaches, exhibiting lower values for all predictive metrics and empirical coverage close to the nominal value under both the $p=2$ and $p=3$ simulation settings.
In order to better understand these results, Figure~\ref{fig:contour_kde2d_scenario_2} displays the posterior distributions obtained by each method for two representative samples from a single replicate under the simulation setting with $p=2$.
The posterior distribution produced by ESS is clearly driven by the elliptical geometry induced by the underlying bivariate Gaussian prior, resulting in dispersed distributions with a considerable amount of draws lying outside the modern climate convex hull.
This explains why ESS consistently overestimates the nominal coverage.
Although cESS incorporates a smooth truncated bivariate normal prior that reduces the number of posterior draws outside the convex hull, hence underestimating the nominal coverage, its posterior distribution retains the same elliptical Gaussian shape.
In contrast, SIR assigns most of its posterior mass near the true climate value, while producing relatively few draws far from it, with all posterior draws fully restricted to the modern climate convex hull.
This behaviour is expected as the underlying uniform prior assigns the same probability to all admissible values within the convex hull.
Consequently, the posterior distribution in \eqref{eq:marginal_cut_posterior} is driven entirely by the ZANIM-LN-BART likelihood, $p(\mathbf{y}_s \mid \mathbf{x}_s, \bm{\Psi})$, such that the inference on $\mathbf{x}_s$ primarily reflects the information from the complex pollen-climate relationships learned in the forward module and the observed fossil counts $\mathbf{y}_s$, rather than the elliptical geometry underpinning the (truncated) Gaussian prior under the (c)ESS schemes.

\begin{table}[!htb]
\centering
\caption{Comparison of the predictive performance of the three sampling schemes proposed in Section~\ref{sec:cut_posterior} to approximate the inverse posterior in simulated settings with $p=2$ and $p=3$ climate covariates.
The metrics are based on $n_f=100$ fossil samples, and the results are averaged across six independent replicates.
The MSEP (MAE) assesses the accuracy of the posterior-predictive mean (median), whereas the ES is a proper scoring rule that evaluates the full multivariate posterior-predictive distribution.
The $95\%$ coverage is calculated based on the $95\%$ HPD region of the two- $(p=2)$ or three- $(p=3)$ dimensional posterior draws, $\{\mathbf{x}^{(r)}_s\}_{r=1}^R$.
Smaller values of ES, MSEP, and MAE indicate better predictive performance, while coverage values close to the nominal level ($0.95$) indicate well-calibrated posterior uncertainty.
Values in bold indicate the best performing method under each predictive metric.}
\label{tab:results_scenario2}
\begin{tabular}{rrrrrrl}
  \hline
 & Sampling scheme & ES & MSEP & MAE  & $95\%$ coverage & Time \\
  \hline
\multirow{3}{*}{$p=2$}&SIR & \textbf{0.692} & \textbf{1.503} & \textbf{1.270} &\textbf{0.932} & $1\mathrm{h}~18\mathrm{min}$ \\
  &ESS & 0.803 & 1.744 & 1.433 & 0.987 & $1\mathrm{h}~1\mathrm{min}$ \\
  &cESS & 0.811 & 1.820 & 1.476 & 0.917 & $1\mathrm{h}~6\mathrm{min}$ \\
\hdashline
\multirow{3}{*}{$p=3$} &SIR & \textbf{0.971} & \textbf{2.512} & \textbf{2.122} & \textbf{0.955} & $1\mathrm{h}~18\mathrm{min}$ \\
  &ESS & 1.181 & 3.354 & 2.539 & 0.995 & $1\mathrm{h}~3\mathrm{min}$ \\
  &cESS & 1.187 & 3.407 & 2.574 & 0.872 & $1\mathrm{h}~10\mathrm{min}$ \\
   \hline
\end{tabular}
\end{table}

The computational cost to reconstruct the $n_f=100$ fossil samples is also reported in Table~\ref{tab:results_scenario2}.
The ESS is the fastest algorithm, with its runtime taking around an hour in both settings.
This is followed by cESS, which has a moderate increase in the runtime, since it spends more time in `slice' steps due to the multivariate truncated normal prior.
As the climate dimension increases from $p=2$ to $p=3$, both the ESS and cESS methods have an increase in their runtime, as they require sampling from higher-dimensional (truncated) normal distributions.
Although SIR is the slowest method, requiring approximately 1 hour and 18 minutes, its computational cost does not increase with the climate dimension $p$.
This is because our SIR approach reuses the same set of $K$ proposal values, and consequently the evaluations of the corresponding regression tree ensembles at the proposal values, across all posterior draws for all $n_f=100$ fossil samples.
Specifically, as described in Section~\ref{sec:sir}, our SIR implementation first generates $K=2{,}000$ proposal climate values and computes the corresponding regression tree ensembles for each of the $R=5{,}000$ posterior draws, $\bm{\Psi}^{(r)}$.
These predictions are stored and subsequently reused during the importance sampling step, eliminating the need to regenerate proposal values and/or compute the predictions of the regression tree ensembles separately for each fossil sample.
Consequently, the total runtime of our SIR approach can be decomposed into two sequential components: (i) generating the uniform proposal climate values and computing the corresponding predictions, and (ii) performing the importance sampling calculations for each fossil sample.
Across both simulation settings, the first stage required approximately 24 minutes, while the latter required approximately 54 minutes.

\begin{figure}[!htb]
    \centering
    \includegraphics[width=1.0\linewidth]{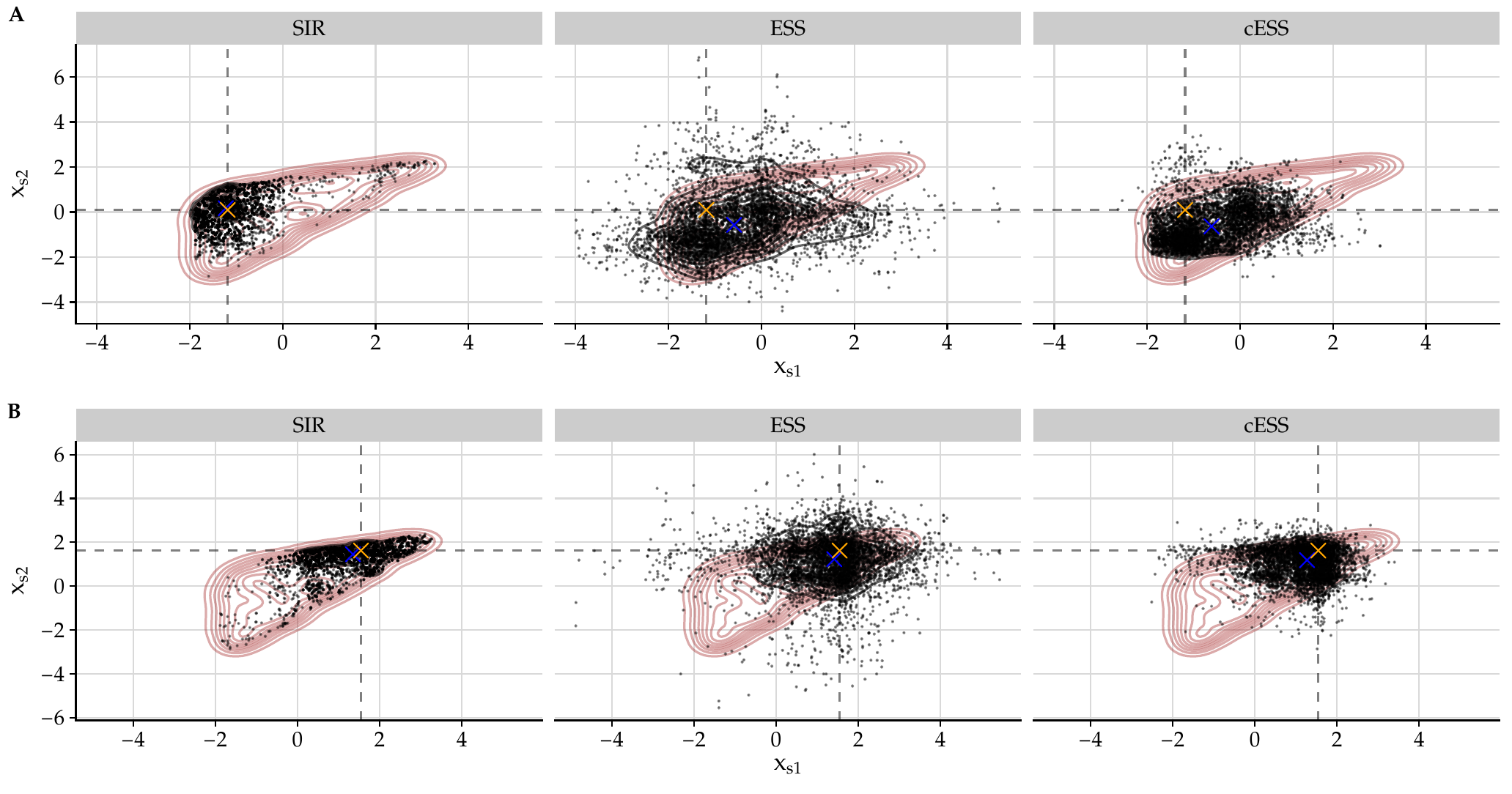}
\caption{Comparison of the obtained inverse posterior distribution of $\mathbf{x}_s=(x_{s1}, x_{s2})$ using SIR, ESS, and cESS  for two representative fossil samples in a given replicate under the simulation setting with $p=2$.
    Each panel displays the posterior draws $\mathbf{x}^{(r)}_s=(x^{(r)}_{s1}, x^{(r)}_{s2})$ as small grey dots, the corresponding kernel density estimates as black contour lines, the posterior medians, indicated by blue cross-marks, and the underlying kernel density estimates of the uniform prior over the convex-hull region $\mathcal{C}$ of the modern climate as light brown contour lines.
    In each case, the true climate values $\mathbf{x}_s$ (Panel \textbf{A}: $\mathbf{x}_s=(-1.184, 0.091)$ and Panel \textbf{B}: $\mathbf{x}_s=(1.545, 1.622)$) are indicated by the intersection of the two dashed lines and orange cross-marks.}
    \label{fig:contour_kde2d_scenario_2}
\end{figure}
\section{Applications}\label{sec:application}

We now apply our methodology to the motivating data sets discussed in Section~\ref{sec:data}.
In Section~\ref{sec:app_forward_fit}, we investigate the fit of our ZANIM-LN-BART model to the full modern calibration data set, which consists of the compositional pollen counts and climate variables collected around the Northern Hemisphere, and assess the estimated pollen-climate relationships from the forward module only.
In Section~\ref{sec:app_validation}, we conduct an experiment to validate the broader reconstruction framework, using a train/test split on the modern data set only, to evaluate our method in terms of reconstructing an out-of-sample three-dimensional modern climate vector composed of the GDD5, MTCO, and AET/PET variables.
Then, in Section~\ref{sec:app_monticchio}, we incorporate the observed fossil pollen counts, along with the modern data, to address our main goal of reconstructing the unobserved past values of the climate covariates at the upland LGdM site in southern Italy.
Unlike the simulation studies where we generated $K=2{,}000$ proposals for the SIR algorithm, we use $K=4{,}000$ for the reconstructions in Section~\ref{sec:app_validation} and \ref{sec:app_monticchio} to ensure adequate exploration of the cut posterior target.

\subsection{Forward module: Estimated pollen-climate relationships from modern data}\label{sec:app_forward_fit}

The goodness-of-fit of the ZANIM-LN-BART model for the modern data set has already been investigated by \citet{Menezes2026},
who showed that it successfully captured the main characteristics of pollen-climate data, including zero-inflation, overdispersion, and complex dependence structures, as well as nonlinearities and interactions in the climate covariate effects.
Moreover, it provided the best fit among a range of competing models across several performance metrics.
The model fit by \citet{Menezes2026} adopted the same hyperparameter settings as the simulation studies in Section~\ref{sec:sim_studies}.
In particular, $M_{\mathrm{c}}=M_0=100$ trees were used, per taxon, for the count and structural zero components, respectively, and the MCMC algorithm was run for $10{,}000$ iterations, with the first $5{,}000$ discarded as burn-in.
Under these settings, the model took approximately $3.4$ hours to run.

In Section~\ref{sec:app_monticchio}, this fitted model will be treated as part of the forward module of our full palaeoclimate reconstruction framework.
This is facilitated by the modular nature of our approach.
For now, we focus on the important inferential outputs of the ZANIM-LN-BART model itself, in the form of three key parameters of interest.
Specifically, we consider the model predictions of the population-level compositional probabilities, $\theta_{ij}=f_j^{(\mathrm{c})}(\mathbf{x}_i)/\sum_{k=1}^df_k^{(\mathrm{c})}(\mathbf{x}_i)$, population-level structural zero probabilities, $\zeta_{ij}=\Phi(f_j^{(0)}(\mathbf{x}_i))$, and individual-level compositional probabilities, $\vartheta_{ij}$, which combine with $\theta_{ij}$, $z_{ij}$, and the logistic-normal random effects via \eqref{eq:zanim_ln_bart_sr}.
The posterior distributions of these quantities provide information on how the underlying compositional probabilities, structural zero probabilities, and abundances of each taxon are affected by the three climate covariates, thereby providing a clear probabilistic description of the pollen-climate relationships uncovered by our model.

Figure~\ref{fig:climate_predictions_chosen} shows heatmaps of the posterior median predictions of these three quantities for three representative taxa.
Each panel of Figure~\ref{fig:climate_predictions_chosen} corresponds to a single taxon and a single pair of climate variables.
Although pairs of covariates are shown, the predictions were computed over an irregular three-dimensional discrete grid of the climate variables (GDD5, MTCO, and AET/PET) representing the modern climate sites at which pollen was actually observed, adapting the approach of \citet{Haslett2006} for three dimensions.
This yields $3{,}120$ grid points, which serve as out-of-sample observations, where each of the three covariates have $31$ unique values.
Note that this grid restricts the modern climate space differently from the convex hull which acts as both the prior and the proposal under the SIR algorithm described in Section~\ref{sec:sir}, or as a constraint on the cESS algorithm described in Section~\ref{sec:ess}.

From Figure~\ref{fig:climate_predictions_chosen}, we can see clear illustrations of how the climate conditions interact to impact the presence, absence, and abundance of the three example taxa.
From Panel \textbf{A}, which plots GDD5 vs. AET/PET and gives predictions for the \textit{Pinus.D} taxon, we see generally high levels of abundance across almost the entire modern climate space, with a notable peak where the GDD5 values are between $1{,}000$ and $2{,}000$ and the AET/PET values are moderate, ranging from $0.30$ to $0.70$.
The largest structural zero probabilities of around $0.25$ occur where the GDD5 values are exactly $0$ and the AET/PET values are below or equal to $0.25$.
Otherwise, the structural zero probabilities remain quite low for this taxon for all but the lowest values of GDD5.
Conversely, from Panel \textbf{B}, which plots MTCO vs. AET/PET and gives predictions for the \textit{Betula} taxon, we see a region of climate space where the taxon is estimated to be entirely absent.
This region occurs where MTCO is at its highest (between $-10^\circ\mathrm{C}$ and $20^\circ\mathrm{C}$) and AET/PET is generally low.
In this region, it is notable that $\theta_{ij}$ is low, $\zeta_{ij}$ is high, and these quantities combine to produce posterior medians of exactly $0$ for the abundances, $\vartheta_{ij}$.
Coupled with the fact that the compositional probabilities are relatively higher and the structural zero probabilities are relatively lower at higher values of AET/PET, across most of the range of MTCO values, this suggests that the \textit{Betula} taxon does not thrive in warmer environments where moisture is scarce.
Finally, from Panel \textbf{C}, which plots GDD5 and MTCO and gives predictions for the \textit{Juniperus} taxon, we see a general pattern of the abundances increasing, and the probability of absence decreasing, as both climate covariates increase in value.
This suggests that this taxon thrives in warmer climates.
This is borne out by the taxon exhibiting non-zero posterior mean estimates of $\vartheta_{ij}$ in only a narrow strip of the observed climate space, where GDD5 exceeds around $4{,}000$ and MTCO is positive.
For all three displayed taxa, their general behaviour across climate space is consistent with the abundances and the dispersion and zero-inflation indices reported for the modern data in Table~\ref{tab:stats_modern_fossil_pollen}, with \textit{Pinus.D} being highly abundant and dispersed, with few zeros, \textit{Betula} being similar with regards to abundance and dispersion, but exhibiting a higher degree of zero-inflation, and \textit{Juniperus} being much less abundant, with a far higher degree of excess zeros.

\begin{figure}[!htb]
    \centering
    \includegraphics[width=1.0\linewidth]{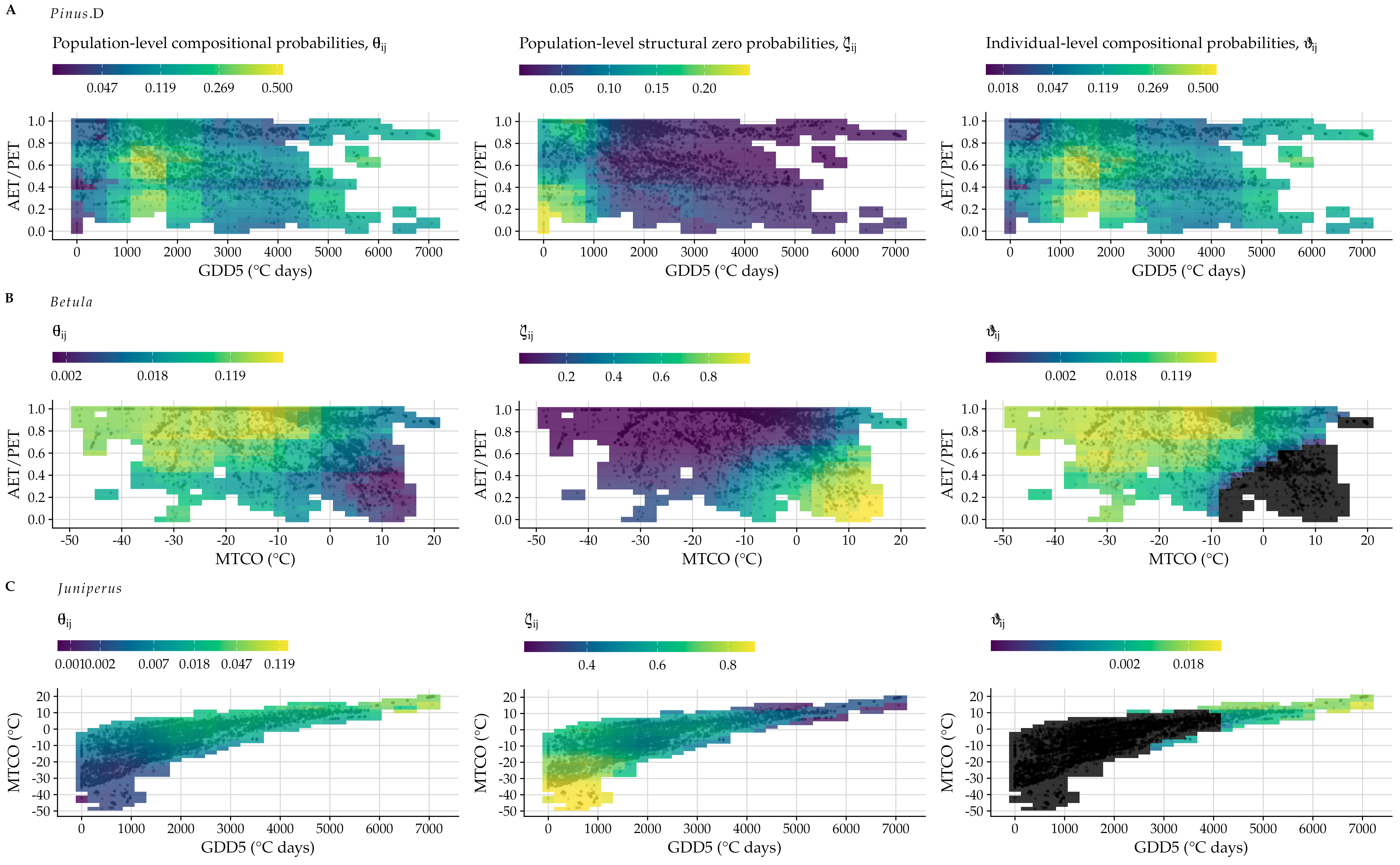}
    \caption{Heatmaps of the posterior median predictions of the ZANIM-LN-BART model parameters for three representative taxa, with the population-level count probabilities ($\theta_{ij}$, left column), population-level structural zero probabilities ($\zeta_{ij}$, middle column) and individual-level count probabilities ($\vartheta_{ij}$, right column).
    The predictions were computed over a three-dimensional discrete grid of the climate covariates (GDD5, MTCO, and AET/PET)
    representing the modern climate sites at which pollen was actually observed.
    Black dots indicate the modern climate observations.
    Each panel represents a distinct taxon and pair of climate covariates, with each bivariate plot showing the model predictions for a given parameter.
    Darker colours correspond to lower parameter values, with the black shading in the right column, showing $\vartheta_{ij}$, corresponding to posterior medians of exactly zero.
    Panel \textbf{A}: \textit{Pinus.D} (GDD5 vs. AET/PET).
    Panel \textbf{B}: \textit{Betula} (MTCO vs. AET/PET).
    Panel \textbf{C}: \textit{Juniperus} (GDD5 vs. MTCO).
    }
    \label{fig:climate_predictions_chosen}
\end{figure}

Although the relationships implied by Figure~\ref{fig:climate_predictions_chosen} are characterised by different, taxon-specific patterns of abundances, structural zeros, and interacting climate covariates, we stress that the three-dimensional climate space should be considered for a more complete picture of the uncovered pollen-climate relationships.
To that end, Appendix~\ref{supp:extra_plots}
provides similar plots, for the same three quantities and for the same three taxa, but across all three pairs of climate covariates.Overall, however, the capabilities of the ZANIM-LN-BART to account for relationships of such complexity as part of the forward module will lead to more accurate palaeoclimate reconstructions, with better uncertainty calibration.
We henceforth incorporate the reconstruction module of our framework using posterior draws of the parameters of ZANIM-LN-BART fits.

\subsection{Validation experiment: Reconstructing holdout modern climate variables}\label{sec:app_validation}

We now assess the performance of our full palaeoclimate reconstruction framework using only the modern data set, for which both the pollen counts and the climate covariates are observed.
Specifically, we randomly select $n_f=1{,}000$ samples to serve as an `out-of-sample' fossil data set, $\{\mathbf{y}_s, \mathbf{x}_s\}_{s=1}^{n_f}$, where the true climate $\mathbf{x}_s$ is treated as known for validation purposes.
The remaining $n_m=6{,}832$ observations form the modern data set, $\{\mathbf{y}_i^m, \mathbf{x}_i^m\}_{i=1}^{n_m}$, which is used to learn the pollen-climate relationships under the forward module using the ZANIM-LN-BART model described in Section~\ref{sec:forward_module}.
Given the train/test split employed for validation purposes, we stress that the ZANIM-LN-BART model fit described in Section~\ref{sec:app_forward_fit} cannot be used.
The runtime for refitting the model to the reduced set of $n_m=6{,}832$ modern observations, using the same hyperparameter and MCMC settings, was approximately $3.2$ hours.
Posterior climate reconstructions are then obtained for each of the $n_f=1{,}000$ holdout samples using the reconstruction module described in Section~\ref{sec:reconstruction_module}.
The resulting posterior samples, $\{\bm{\Psi}^{(r)}\}_{r=1}^{R}$, with $R=5{,}000$, were subsequently used in the reconstruction module for each holdout fossil sample.
As stated, our focus here is purely on evaluating the reconstruction module.
Specifically, we compare the three sampling schemes introduced in Section~\ref{sec:cut_posterior} (SIR, ESS, and cESS) using the same performance metrics adopted in the simulation studies in Section~\ref{sec:sim_studies}.
These metrics compare the known climate vector $\mathbf{x}_s$ of each holdout fossil sample with its corresponding inverse posterior distribution.

The results are presented in Table~\ref{tab:oos_pollen_climate}.
The findings closely mirror those of the simulation study in Section~\ref{sec:sim_studies}, which was designed to match the present application in some respects.
Among the three sampling schemes, SIR consistently provides the best predictive performance, achieving the lowest ES, MSEP, and MAE, while also producing an empirical joint coverage ($0.9510$) quite close to the nominal $95\%$ level.
These results indicate that SIR not only produces the most accurate point reconstructions but also yields well-calibrated posterior uncertainty.
In contrast to the simulation studies of Section~\ref{sec:sim_studies} with $p=3$, here we can see more appreciable improvements of the cESS scheme over the ESS scheme.
The ESS algorithm overestimates the empirical coverage ($0.9860$) while the empirical coverage of cESS ($0.9480$) is much closer to the nominal value.
The cESS approach also shows a modest improvement in the ES metric relative to the ESS algorithm.
These improvements are likely due to the truncated normal prior employed by cESS, which smoothly constrains the posterior samples to the convex hull of the observed modern climate and therefore limits implausible climate reconstructions.
Our SIR approach, however, avoids potentially implausible reconstructions entirely, by virtue of its use of a continuous uniform prior over the convex hull of the modern climate.
This fact may explain its superiority across all reported metrics.

\begin{table}[!htb]
\caption{Comparison of the predictive performance of the three sampling schemes proposed in Section~\ref{sec:cut_posterior} to predict the holdout three-dimensional modern climate vector: GDD5, MTCO, and AET/PET.
We use a subset of $n_m=6{,}832$ samples from the modern data to fit the ZANIM-LN-BART model as part of the forward module, then assess the performance of the methods in terms of reconstruction based on the remaining $n_f=1{,}000$ samples considered as the fossil data set.
The MSEP (MAE) assesses the accuracy of the posterior-predictive mean (median), whereas the ES is a proper scoring rule that evaluates the full multivariate posterior-predictive distribution.
The $95\%$ coverage is calculated based on the $95\%$ HPD region of the three-dimensional posterior draws, $\{\mathbf{x}^{(r)}_s\}_{r=1}^R$.
Smaller values of ES, MSEP, and MAE indicate better predictive performance, while coverage values close to the nominal level ($0.95$) indicate well-calibrated posterior uncertainty.
Values in bold indicate the best performing method under each predictive metric.}
\centering
\label{tab:oos_pollen_climate}
\begin{tabular}{lrrrrl}
  \hline
Sampling scheme & ES & MSEP & MAE  & $95\%$ coverage & Time \\
  \hline
  SIR & \textbf{0.721} & \textbf{1.530} & \textbf{1.483} & \textbf{0.951} & $14\mathrm{h}~57\mathrm{min}$ \\
  ESS & 0.988 & 2.413 & 2.009 & 0.986 & $10\mathrm{h}~6\mathrm{min}$ \\
  cESS & 0.954 & 2.527 & 1.986 & 0.948 & $11\mathrm{h}~10\mathrm{min}$ \\
\hline
\end{tabular}
\end{table}

Regarding the computational cost in reconstructing the $n_f=1{,}000$ holdout fossil samples, SIR has the largest runtime, followed by cESS and ESS, requiring approximately $15$, $11$, and $10$ hours, respectively.
As discussed in Section~\ref{sec:sir}, the runtime of our SIR approach can be decomposed into two sequential tasks.
The pre-computations in the first stage, which is performed only once, required $38$ minutes.
Following this, the importance weighting and resampling steps in the second stage required an additional $14\mathrm{h}~19\mathrm{min}$.
Consequently, although the total runtime is higher than for (c)ESS, the average computational time taken to reconstruct a single holdout fossil sample after performing the necessary pre-computations is only $51.5$ seconds.
Overall, this validation experiment is consistent with the simulation studies in providing empirical evidence that the SIR sampling scheme is the preferred reconstruction algorithm for the proposed framework, as it consistently delivers the most accurate reconstructions while maintaining appropriate uncertainty quantification.

\subsection{Reconstruction module: A case study from Lago Grande di Monticchio}\label{sec:app_monticchio}

Next, we apply our methodology with the goal of reconstructing the three past climate variables, GDD5, MTCO, and AET/PET, for the case study of LGdM described in Section~\ref{sec:data}.
We employ the fitted ZANIM-LN-BART model from Section~\ref{sec:app_forward_fit}, which used all $n_m=7{,}832$ samples from the modern data set.
We stress that, due to the modularisation of our framework arising from the cut posterior approach, this model fitting needs to be performed only once.
Coupled with our use of a large modern calibration data set, this implies that palaeoclimate reconstructions, at any number of sites for which corresponding fossil pollen counts are available for the same taxa in the modern data, can then be performed using the posterior draws of the parameters of a single fitted ZANIM-LN-BART model.

Subsequent to fitting the ZANIM-LN-BART model to the modern calibration data, we use the $R=5{,}000$ posterior samples of the model parameters, $\{\bm{\Psi}^{(r)}\}_{r=1}^{R}$, in the reconstruction module to propagate uncertainty and compute the inverse posterior distribution of the unobserved past climate $\mathbf{x}_s$ using the pollen counts $\mathbf{y}_s$ for all $n_f=924$ fossil samples.
Following its superior performance in the validation experiments of Section~\ref{sec:app_validation}, we use our SIR approach to approximate the inverse posterior distribution in the reconstruction module.
The total runtime to reconstruct the climate variables across all $n_f=924$ fossil samples was $14\mathrm{h}~24\mathrm{min}$,
with the first stage requiring $38\mathrm{min}$ to compute the predictions for the regression tree ensembles, and the second stage requiring an additional $13\mathrm{h}~46\mathrm{min}$
to perform the importance weighting and resampling steps for each fossil sample.
Consequently, the total computational cost to reconstruct a single fossil sample after performing the necessary pre-computations is around $53$ seconds, on average.

We recall that each fossil sample $s\in\{1,\ldots, n_f\}$ has an associated age in thousands of years before present, where `present' corresponds to 1950 AD.
We follow the scientific dating literature by using the term `calibrated' to indicate that these ages are expressed on a calendar-year timescale.
Thus, we use the unit ka BP as shorthand for thousands of calibrated years before present.
We use this information to present the results in Figure~\ref{fig:monticchio_reconstruction}, although we emphasise that our methods do not incorporate this information directly.
The panels in Figure~\ref{fig:monticchio_reconstruction} present the posterior median and the $95\%$ credible interval for each climate variable along the ages associated to each fossil sample.
Gaps in the chronology occur at approximately 27.7--30.3 ka BP and 38.7--40.1 ka BP.
Consequently, our reconstructions provide straight line predictions during these periods.

Although the reconstructed climate variables exhibit variability, the reconstruction reveals several coherent long-term changes in the LGdM climate.
One prominent feature is the interval between approximately 20--14 ka BP, during which MTCO and AET/PET both increase markedly, indicating progressively warmer winters and increasing moisture availability.
These changes typify this interval of transition following the Last Glacial Maximum (LGM), which spanned approximately 26--20 ka BP, during which both winter temperatures and moisture availability were much reduced compared to the present.
Another pronounced transition occurs between 93--90 ka BP, when GDD5 decreases abruptly while AET/PET increases rapidly.
Taken together, the reconstructed climate vectors suggest alternating intervals characterised by (i) long, warm growing seasons with mild winters and abundant moisture, and (ii) cold, moisture-limited climates with shorter growing seasons.
These shifts are most evident around 15 ka BP and 90 ka BP, where pairs of climate variables change simultaneously.
To provide a closer examination of the effects of the aforementioned shifts, Figure~\ref{fig:kde2d_reconstruction_age_9_23} presents the palaeoclimate reconstructions for the LGdM site for two fossil samples with corresponding ages of 23.950 ka BP and 9.831 ka BP.
These samples were chosen to illustrate the transitions between the LGM and the Holocene.
We summarise the reconstructions for these samples using bivariate plots for the three pairs of climate variables (GDD5, MTCO, and AET/PET), which display the posterior draws, kernel density estimates thereof, and the corresponding posterior medians, along with the underlying kernel density estimates of the uniform prior over the convex-hull region $\mathcal{C}$ which defines the modern climate domain.

\begin{figure}[!htb]
    \centering
    \includegraphics[width=1\linewidth]{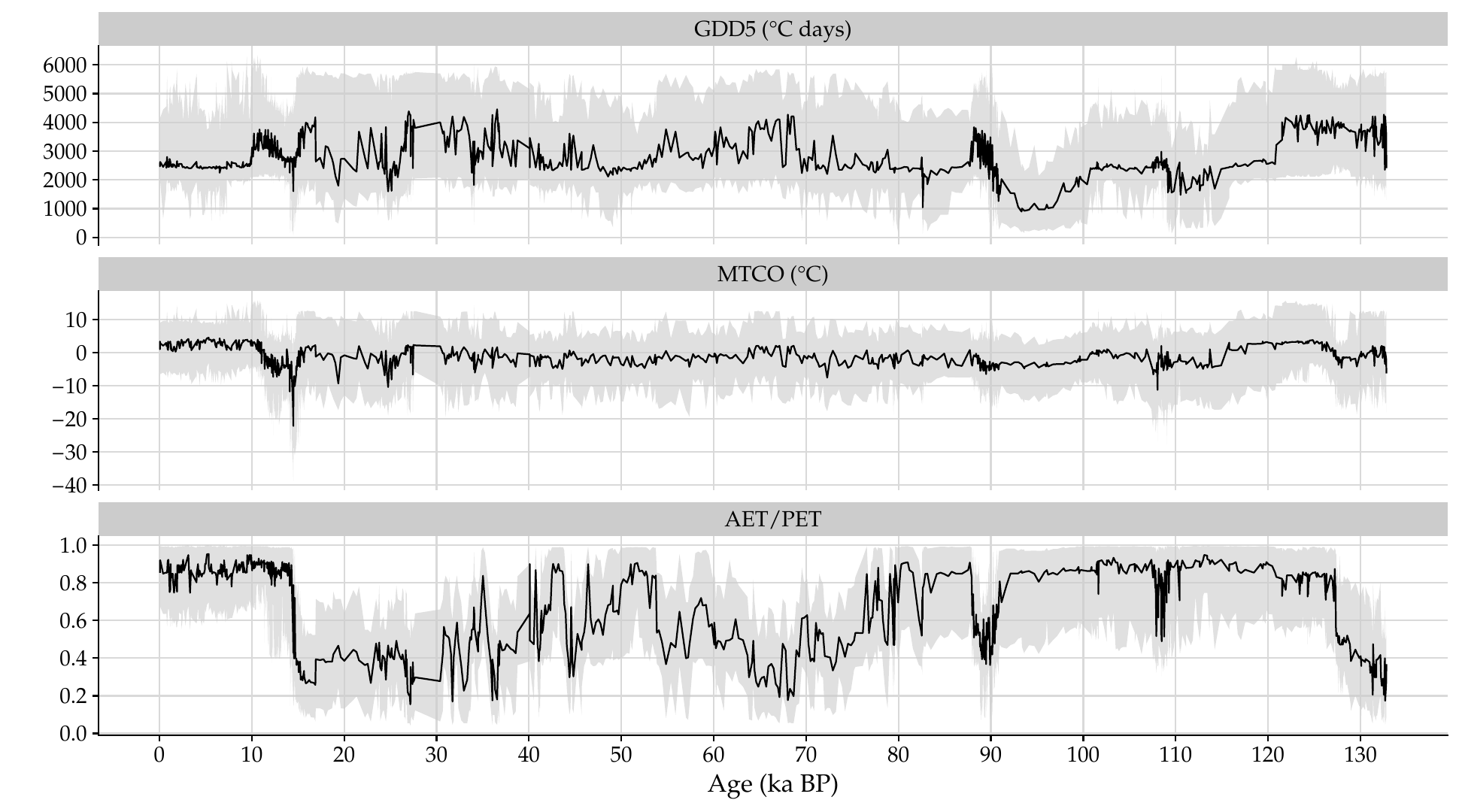}
\caption{
    Results of the palaeoclimate reconstruction using the proposed methodology for the Lago Grande di Monticchio site.
    Each panel presents posterior summaries (median as black lines and $95\%$ credible intervals as grey ribbons) for the reconstruction of the three climate variables, GDD5 (top), MTCO (middle), and AET/PET (bottom), across all $n_f=924$ fossil samples. Each fossil sample has an associated calibrated age in ka BP, which is used to depict the palaeoclimate reconstructions as time series plots, from youngest to oldest.}
    \label{fig:monticchio_reconstruction}
\end{figure}

From Panel \textbf{A} of Figure~\ref{fig:kde2d_reconstruction_age_9_23}, which shows the reconstruction of the fossil samples within the LGM period, we have clear evidence of two distinct modes for all the three bivariate plots.
It is notable from the plot of GDD5 vs. AET/PET, in particular, that our sampling scheme is capable of traversing somewhat disconnected modes in the inverse posterior.
The most dominant mode for the bivariate plot of GDD5 vs MTCO is located at an MTCO value around $-10^{\circ}$ and a GDD5 value near $2{,}000$.
These values are consistent with previous findings concerning the LGM period in this upland Italian region \citep{Watts1996,Parnell2016}.
Although the value of MTCO at the most dominant mode is below $0^\circ\mathrm{C}$, this is less extreme than the coldness experienced in other regions throughout the LGM.
Indeed, Mediterranean climates enjoyed less severe winter harshness than glaciated regions and ice-sheet zones in central and northern parts of Europe during this period.
Furthermore, the GDD5 value of around $2{,}000$ at the most dominant mode is indicative of the depressed summer temperatures in southern Europe during the LGM.
Finally, the modal AET/PET value evident from the most dominant modes in the other two plots in Panel \textbf{A} of Figure~\ref{fig:kde2d_reconstruction_age_9_23}, around $0.39$, reflects the limited precipitation in this arid region during the LGM.

Conversely, in the reconstructions in Panel \textbf{B} of Figure~\ref{fig:kde2d_reconstruction_age_9_23}, over ten thousand years after the LGM, there is evident warming of the winter months, as the values of MTCO have substantially increased, relative to Panel \textbf{A}.
Notably, the modal value of MTCO is now above $0^\circ\mathrm{C}$.
The increase in moisture availability is also apparent as the posterior distribution of AET/PET is generally greater than $0.7$ and almost all posterior mass has shifted into regions of AET/PET that were not explored by the sampler in the reconstructions at age 23.905 ka BP.
Moreover, these reconstructions at 9.831 ka BP exhibit tighter concentrations around a single mode in each bivariate illustration.
This is particularly evident in the plots involving GDD5, where the mode around $2{,}000$, which was formerly the most dominant of two modes, is now the only mode, with most of the posterior samples clustered tightly around this region of climate space.
Indeed, there is evident skewness in the posterior distributions in the plots involving GDD5 at 9.831 ka BP.

Notably, the reconstructed climate at 9.831 ka BP is more similar to modern climatic conditions at the LGdM site than the reconstructions at 23.905 ka BP.
The median values of the reconstructions of the three climate variables nearest the present day (i.e., the most recent fossil sample at 0.0587 ka BP) are 2598.9 (GDD5), 3.27$^\circ\mathrm{C}$ (MTCO), and 0.85 (AET/PET), while the respective medians of the reconstructions at 9.831 ka BP are 2521.787, 3.41$^\circ\mathrm{C}$ , and 0.94.
These similarities are also evident in Figure~\ref{fig:monticchio_reconstruction}, which shows reasonable stability across all three climate variables from around 10 ka BP to the most recent records, near the present day.
However, it is also interesting to compare the most recent reconstructions at 0.0587 ka BP to the values of the climate covariates of the four samples collected at LGdM in our modern data set, which range from 2859 to 2886 (GDD5), 3.9$^\circ\mathrm{C}$ to 4.2$^\circ$ (MTCO), and 0.664 to 0.707 (AET/PET).
This suggests that considerable warming has taken place at this site in the intervening years between the most recent fossil records and the present day.

\begin{figure}[!htb]
    \centering
    \includegraphics[width=1.0\linewidth]{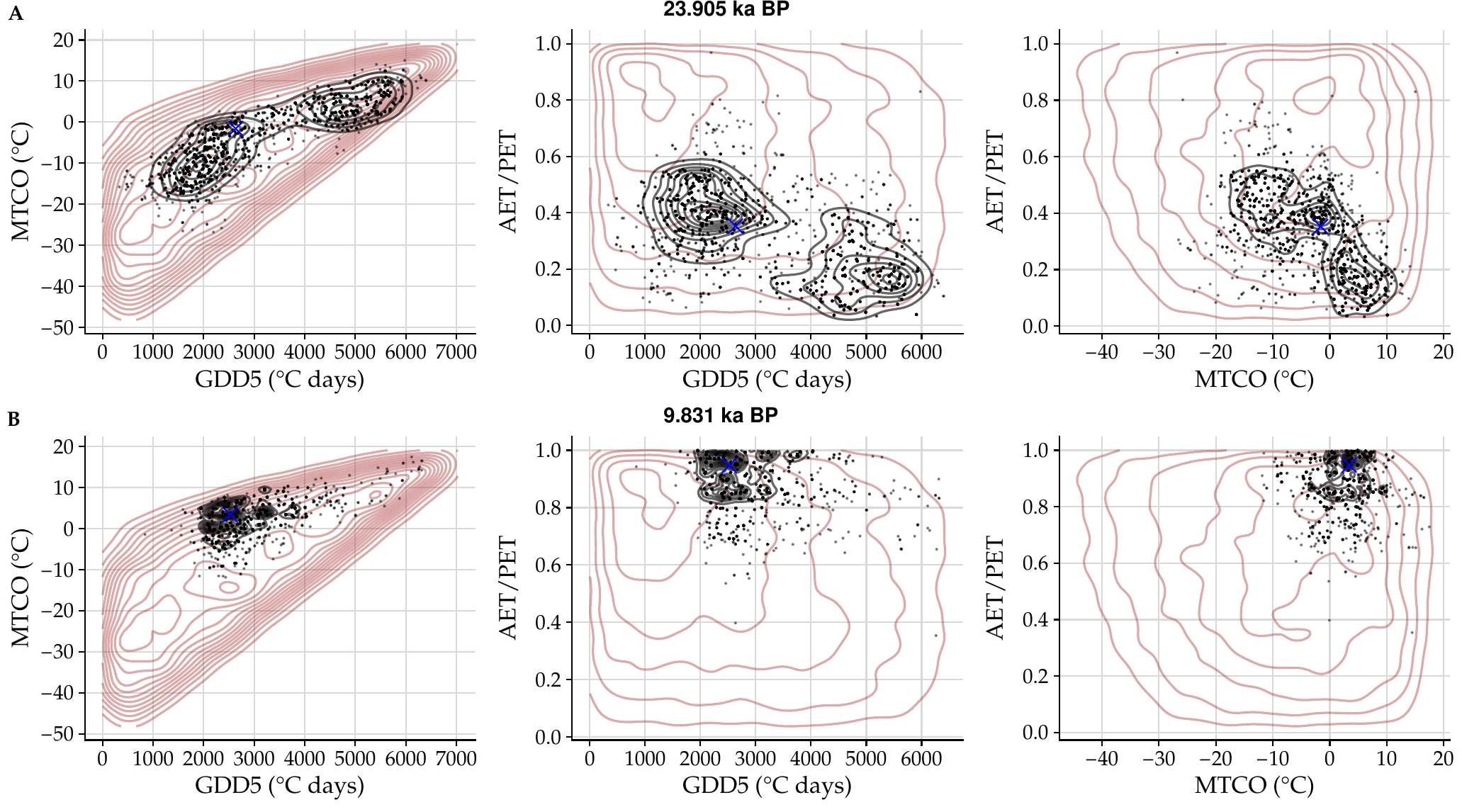}
\caption{Palaeoclimate reconstructions of two fossil samples with ages of 23.950 ka BP (Panel \textbf{A}) and 9.831 ka BP (Panel \textbf{B}) for the Lago Grande di Monticchio site.
    For the three pairs of climate variables, each panel displays the posterior draws as small grey dots, the two-dimensional kernel density estimates thereof as black contour lines, the posterior medians, indicated by the blue cross-mark, and the underlying kernel density estimates of the uniform prior over the convex-hull region $\mathcal{C}$ of the modern climate as light brown contour lines.}
    \label{fig:kde2d_reconstruction_age_9_23}
\end{figure}
\section{Discussion}\label{sec:discussion}

We present a new probabilistic modular framework to perform palaeoclimate reconstruction from fossil pollen records that accommodates structural zeros, allows flexible taxon-specific responses to multiple climate variables, and provides coherent uncertainty quantification.
We formulate our modular framework within the cut posterior approach \citep{Plummer2015}, comprising two modules: a forward module and a reconstruction module.
The ZANIM-LN-BART model which we adopt to describe the pollen-climate relationships governs both modules of our framework.
This model coherently specifies a zero-inflated count-compositional probability distribution for the pollen counts, accommodating structural zeros, overdispersion, and complex dependence among taxa.
Furthermore, the model flexibly captures nonlinear effects and interactions between climate covariates in both the compositional and structural zero components.
This represents a key contribution in that existing methodologies address some but not all of these inherent issues with the data.

Within the forward module, we use a modern calibration data set from the Northern Hemisphere to perform inference on the parameters of the ZANIM-LN-BART model, hence the information from the fossil data is intentionally cut in the updates of the model parameters.
In the reconstruction module, we propagate the uncertainty from the forward module to perform inference on the missing past climate values.
We propose and evaluate three sampling schemes for doing so.
Our primary approach combines the sampling importance resampling (SIR) algorithm with the multiple imputation strategy of \citet{Plummer2015}.
This algorithm employs a continuous uniform prior constrained to the convex hull of the modern climate.
Notably, our prior assigns mass over the entire modern climate domain, whereas alternative methods which rely on discretisation to specify a prior which similarly encompasses the modern climate space \citep[e.g.,][]{SalterTownshend2012,Parnell2015} require a computationally onerous increase in the number of grid-points to approximate this level of granularity.
The effects of our prior are evident in our reconstructions for the LGdM site in our case study in Section~\ref{sec:app_monticchio}, where the reconstructed trajectories are reasonably smooth.
Conversely, the reconstructions for the same site obtained by \citet{Parnell2015} through the discretisation approach of \citet{SalterTownshend2012} exhibit a considerable number of spikes, particularly for the the AET/PET variable.
The convex hull was feasible to compute in our case, given that we aimed to reconstruct three-dimensional climate and the number of modern data observations was not too large, but this strategy may suffer in reconstruction problems that are higher-dimensional or involve a significantly larger number of modern observations \citep{Barber1996}.
In addition, we implemented an elliptical slice sampling (ESS) algorithm \citep{Murray2010}, which assumes a multivariate normal prior for the past climate variables, calibrated using the empirical moments of the modern climate data.
We further proposed a constrained ESS (cESS) algorithm based on a truncated multivariate normal prior that restricts the support to the observed modern climate domain, analogous to the SIR approach, and enforces the constraint through a smooth approximation technique \citep{Maatouk2026}.

Our simulation studies in Section~\ref{sec:sim_studies} and Appendix~\ref{supp:add_sim} demonstrated the importance of explicitly accounting for structural zeros in the reconstruction of past climates.Methods that ignore structural zeros, namely the DM-BUMMER \citep{Vasko2000} and DM-GP \citep{Tipton2019} models, showed poorer predictive performance and lower empirical coverage than the proposed ZANIM-LN-BART model, which produced more accurate reconstructions with well-calibrated uncertainty.
Regarding the comparison of sampling schemes for the reconstruction module, the proposed cESS algorithm outperformed ESS and SIR when a single climate covariate ($p=1$) was considered, owing to its truncated Gaussian prior.
However, as the climate dimension increased ($p=2$ and $p=3$), SIR consistently outperformed both ESS and cESS.
We investigated the poor performance of the ESS and cESS algorithms in settings with $p\geq2$ in Section~\ref{sec:sim_studies} and concluded, based on the clearly elliptical geometry of the resulting inverse posterior distributions in Figure~\ref{fig:contour_kde2d_scenario_2}, that the (truncated) multivariate Gaussian prior underlying the (c)ESS algorithm plays a fundamental role in the obtained palaeoclimate reconstructions.
In contrast, the ZANIM-LN-BART likelihood terms appear less susceptible to being overwhelmed by our continuous uniform prior over the domain of the modern climate variables under our SIR approach.

After characterising the pollen-climate relationships uncovered by the ZANIM-LN-BART model applied to the modern calibration data, for which a three-dimensional climate vector (GDD5, MTCO, and AET/PET) is observed, we also performed an empirical comparison of the sampling schemes in a validation experiment using the modern data set and assessed the ability of our framework to reconstruct the holdout climate variables.
Consistent with the simulation studies, the SIR algorithm outperformed both ESS and cESS, providing more accurate reconstructions and well-calibrated empirical coverage.
Finally, we applied the proposed framework, coupled with the SIR algorithm, to reconstruct the three-dimensional climate vector from fossil pollen counts collected at Lago Grande di Monticchio in southern Italy.
The reconstructed posterior median climate trajectories exhibited substantial long-term fluctuations across all three climate variables over the range of ages for which fossil pollen counts were available for the LGdM site.
In particular, the interval between 20 and 14 ka BP showed marked increases in MTCO and AET/PET, indicating warmer winters and increased moisture availability.
These findings indicated substantial changes in the climatic conditions after the Last Glacial Maximum period, which occurred approximately $26{,}000$ to $20{,}000$ years ago.
We also provided bivariate plots summarising the palaeoclimate reconstructions at the LGdM site corresponding to two fossil pollen samples; one observed during the LGM period, and another at around 9 ka BP, i.e., during the early Holocene.
These summaries showed moderately harsh winters, depressed summer temperatures, and limited precipitation during the LGM period, along with substantial warming and the increased availability of moisture thereafter.
These summaries also illustrated that our framework can capture the multimodal climatic conditions present during the LGM period.
Finally, our results showed that the climate at LGdM remained stable from around 9 ka BP to the present day.

Our framework assumes that the unobserved climate covariates are conditionally independent given the fossil counts and the estimated parameters.
We then consider different prior distributions according to the sampling scheme adopted for the reconstruction module: a uniform prior restricted to the convex hull of the modern climate under SIR, and a (truncated) multivariate Gaussian prior under (c)ESS.
However, these assumptions ignore the temporal dependence induced by the ages of the fossil samples, which are themselves continuous variables.
A natural extension of our framework would be to incorporate the uncertainty in the ages of the samples using an age-depth chronology model \citep[e.g.,][]{Haslett2008}.
However, this is less of a concern for the present application, as the age estimates for LGdM are derived from annual lamination counts and tephrachronology, which can provide smaller uncertainties than radiocarbon dating \citep{Brauer2007}.
Another promising extension would be to incorporate temporal dependence through a continuous time Markov process prior for the unobserved climate variables.
Although our reconstructed trajectories for the Lago Grande di Monticchio site are reasonably smooth, we expect that such an extension would both reduce reconstruction uncertainty and produce even smoother trajectories.
When coupled with a Markov process prior, the reconstruction module would naturally admit a state-space representation, in which the filtering and smoothing distributions become the primary inferential objects for palaeoclimate reconstruction \citep{Carson2018}.
However, this extension is non-trivial.
First, the resulting state-space model would involve a multivariate non-Gaussian observation equation with a nonlinear functional form, together with a multivariate latent state (the three-dimensional climate space in our case study).
Second, it would be necessary to specify a state-space prior that incorporates information from the modern climate, analogous to the truncated multivariate Gaussian prior constrained by the convex hull.
Third, this extension would need to be integrated within the cut posterior paradigm to ensure proper propagation of uncertainty from the forward module.
Taken together, these challenges entail substantial computational demands and may motivate the development of new scalable methodological approaches.

In conclusion, we believe that the proposed framework, based on the ZANIM-LN-BART model combined with a modular cut posterior approach, is particularly useful to perform Bayesian palaeoclimate reconstruction from fossil pollen counts using large modern calibration data sets.
Above all, we wish to highlight the crucial importance of modelling the compositional and latent dependencies among the pollen counts and accounting for structural zeros in order to produce accurate, well-calibrated reconstructions.
As our framework is based on the ZANIM-LN-BART model which simultaneously addresses all of these concerns without requiring pre-specification of the functional forms governing the complex pollen-climate relationships, we hope that it will be of practical interest and utility for other researchers aiming to reconstruct past climates at other sites.
Although our interest in the application was solely in the site at LGdM, interested practitioners could, in principle, perform reconstructions for other sites for which corresponding fossil pollen counts are available for the same taxa as in our modern calibration data set, using the same fitted ZANIM-LN-BART model.
Furthermore, the proposed methodology offers a foundation for more sophisticated extensions that incorporate spatio-temporal dependence and age uncertainty.
Although our application considered $p=3$ climate variables, we also note that our framework could be extended to higher-dimensional reconstruction tasks in a reasonably straightforward manner.
To facilitate the adoption and reproducibility of our methods, we provide an open source implementation through the \textsf{R} package \texttt{zanicc} available from \url{https://github.com/AndrMenezes/zanicc}.

\section*{Acknowledgments}
{\small
Andr{\'e} F. B. Menezes's work was supported by Taighde {\'{E}}ireann -- Research Ireland under Grant number 18/CRT/6049.}

{
\interlinepenalty=10000
\setlength{\bibsep}{6.5pt}
\bibliographystyle{agsm}
\bibliography{references}
}

\section*{Appendices}

These Appendices contain: details for the sampling schemes proposed for the reconstruction module in Section~\ref{sec:reconstruction_module} (Appendix~\ref{supp:algorithms}); an additional simulation study comparing our approaches against existing alternatives with software implementations for reconstructing one-dimensional climate (Appendix~\ref{supp:add_sim});  additional heatmaps of the ZANIM-LN-BART model predictions for all pairs of climate variables and taxa, complementing the heatmaps presented in Section~\ref{sec:app_forward_fit} (Appendix~\ref{supp:extra_plots}); and links to \textsf{R} code implementations of the methodology and scripts to reproduce the Lago Grande di Monticchio case study (Appendix~\ref{supp:R_scripts}).

{
\setcounter{section}{0}
\setcounter{subsection}{0}
\setcounter{figure}{0}
\setcounter{table}{0}
\setcounter{equation}{0}
\setcounter{algocf}{0}
\counterwithin{figure}{section}
\counterwithin{table}{section}
\renewcommand{\thetable}{\Alph{section}.\arabic{table}}
\renewcommand{\thefigure}{\Alph{section}.\arabic{figure}}
\renewcommand{\theequation}{\Alph{section}.\arabic{equation}}
\renewcommand{\thealgocf}{\Alph{section}.\arabic{algocf}}
\renewcommand{\AlFnt}{\linespread{0.75}\selectfont}

\renewcommand{\thesection}{\Alph{section}}
\titleformat{\section}
  {\normalfont\Large\bfseries}
  {Appendix~\thesection}
  {1em}
  {}

\section{Inverse posterior sampling algorithms}\label{supp:algorithms}

In Section~\ref{sec:reconstruction_module}, we presented three sampling schemes to approximate the inverse posterior distribution in \eqref{eq:marginal_cut_posterior} of the unknown past climate vector $\mathbf{x}_s$ given the corresponding compositional pollen counts, $\mathbf{y}_s$, in the fossil data and the posterior draws of the ZANIM-LN-BART parameters, $\{\bm{\Psi}^{(r)}\}_{r=1}^R$.
We proposed an approach based on the SIR algorithm in Section~\ref{sec:sir} and presented two other approaches based on the ESS and cESS sampling methods in Section~\ref{sec:ess}.
Here, we present further details on the implementations of these algorithms, with a particular focus on the subroutines required for computing an unbiased and consistent estimate of the ZANIM-LN-BART likelihood, sampling uniformly over the convex hull $\mathcal{C}$ which defines the domain of the modern climate data, and pre-computing the predictions of the model parameters from the regression tree ensembles of the ZANIM-LN-BART model.

We recall that a key challenge, for all three proposed sampling schemes, is that the ZANIM-LN-BART likelihood appearing in the inverse posterior distribution in \eqref{eq:marginal_cut_posterior} is not available in closed-form due to the logistic-normal random effects $\{\mathbf{u}_i\}_{i=1}^n$.
To obtain an unbiased and consistent estimate of the ZANIM-LN-BART likelihood, conditional on any climate covariate values $\mathbf{x}_i$, the regression tree parameters, $\bm{\Psi}^{(\mathrm{c})}$ and $\bm{\Psi}^{(0)}$, and the covariance matrix of the random effects, $\bm{\Sigma}_U$, we exploit the stochastic representation in \eqref{eq:zanim_ln_bart_sr}.
Algorithm~\ref{alg:unbiased_estimate_zanim_ln} describes the resulting Monte Carlo estimator.
It assumes that, for a given vector of climate covariate values $\mathbf{x}_i$, the predictions of the taxon-specific regression tree ensembles for the compositional probabilities, $\theta_{ij}\propto f_j(\mathbf{x}_i; \Psi_j^{(\mathrm{c})})$, and structural zero probabilities, $\zeta_{ij}=\Phi\lbrack f_j(\mathbf{x}_i; \Psi_j^{(0)})\rbrack$, have been computed under the ZANIM-LN-BART model.
In addition, Algorithm~\ref{alg:unbiased_estimate_zanim_ln} also requires $\bm{\Sigma}_U$,
the observed modern compositional counts $\mathbf{y}_i$, and the number of independent Monte Carlo samples $T$.
Algorithm~\ref{alg:unbiased_estimate_zanim_ln} will subsequently be used within both Algorithm~\ref{alg:sir_multiple_imputation} and Algorithm~\ref{alg:ess}, which describe our SIR and (c)ESS approaches, respectively.
Recall from Section~\ref{sec:reconstruction_module} that $T=1$ is used under the SIR approach, while a larger value of $T=100$ is required to obtain a consistent estimate under the (c)ESS approach, though the estimate remains unbiased in each case.

\begin{algorithm}[!htb]
\caption{Compute an unbiased and consistent estimate of the ZANIM-LN-BART likelihood.}
\label{alg:unbiased_estimate_zanim_ln}
\SetAlgoLined
\SetAlgoItemize
\SetKwInput{KwInput}{Input}
\SetKwInput{KwOutput}{Output}
\KwInput{
Number of independent Monte Carlo samples $T$, compositional counts $\mathbf{y}_i$, and the ZANIM-LN-BART parameters, i.e., the compositional probabilities $\bm{\theta}_{i}=(\theta_{i1}, \ldots, \theta_{id})$,
structural zero probabilities $\bm{\zeta}_{i}=(\zeta_{i1}, \ldots, \zeta_{id})$ which are functions of their corresponding taxon-specific regression tree ensembles evaluated at given climate covariate values $\mathbf{x}_i$, and the covariance matrix $\bm{\Sigma}_U$ of the random effects $\mathbf{u}_i$.
}
\KwOutput{An unbiased estimate of the ZANIM-LN-BART likelihood.}

\For{$t$ from $1$ to $T$} {

Sample $z_{ij} \sim \operatorname{Bernoulli}\left\lbrack 1 - \zeta_{ij}\right\rbrack$ if $y_{ij}=0$, otherwise set $z_{ij}=1$.

Sample $\mathbf{u}_i \sim \operatorname{Normal}_d\left\lbrack \mathbf{0}_d, \bm{\Sigma}_U\right\rbrack$.

Compute
\[\vartheta^{(t)}_{ij} = \frac{z_{ij}e^{u_{ij}}\theta_{ij}}{\sum_{k=1}^d z_{ik}e^{u_{ik}}\theta_{ik}}.\]
}

Compute the estimate as follows
\[
\hat{p}(\mathbf{y}_i \mid \bm{\theta}_i, \bm{\zeta}_i, \bm{\Sigma}_U)=
\frac{1}{T}\sum_{t=1}^T \Pr\left\lbrack \mathbf{Y}=\mathbf{y}_i; \bm{\vartheta}^{(t)}_i \right\rbrack,
\]
where $\Pr \lbrack \mathbf{Y} = \mathbf{y}; \bm{\vartheta} \rbrack = \binom{N}{y_{1} \dots y_{d}}\prod_{j=1}^d\vartheta_j^{y_{j}}$
with $N=\sum_{j=1}^d y_j$ is the multinomial PMF.

Return $\hat{p}(\mathbf{y}_i \mid \bm{\theta}_i, \bm{\zeta}_i, \bm{\Sigma}_U)$.
\end{algorithm}

\begin{algorithm}[!htb]
\caption{Generate uniform random points inside the convex hull of $\mathbf{X}$.}
\label{alg:runif_convexhull}
\SetAlgoLined
\SetAlgoItemize
\SetKwInput{KwInput}{Input}
\SetKwInput{KwOutput}{Output}
\KwInput{Number of samples $K$ and  a matrix of observed points $\mathbf{X} \in \mathbb{R}^{n \times p}$, with each row $\mathbf{x}^\top_i = (x_{i1}, \ldots ,x_{ip})$ denoting a point in $\mathbb{R}^p$.
}
\KwOutput{$\{x_k\}_{k=1}^K$, $K$ independent samples uniformly distributed over the convex hull of $\mathbf{X}$.}
Compute the Delaunay triangulation of $\mathbf{X}$:
$\operatorname{DT}(\mathbf{X}) = \{\Delta_1, \ldots, \Delta_L\}$, where each non-overlapping simplex $\Delta_\ell$, $\ell\in\{1,\ldots, L\}$
is defined by $p+1$ rows of $\mathbf{X}$.
The vertices of simplex $\Delta_\ell$ are denoted by a matrix $\mathbf{Z}_\ell = \lbrack\mathbf{x}_{\ell_1}, \ldots, \mathbf{x}_{\ell_{p+1}}\rbrack^\top \in \mathbb{R}^{(p+1) \times p}$.

For each simplex $\Delta_\ell$, $\ell\in\{1,\ldots, L\}$, compute its volume $v_\ell$.

Normalise to obtain the corresponding probabilities $p_\ell = v_\ell / \sum_{h=1}^L v_h$ for each simplex $\ell\in\{1,\ldots, L\}$.

\For{$k$ from $1$ to $K$} {

Generate barycentric coordinates $\mathbf{w} = (w_1, \ldots, w_{p+1}) \sim \operatorname{Dirichlet}\lbrack 1, \ldots, 1\rbrack$.

Draw a simplex index $\ell \in \{1, \ldots, L\}$ according to the probabilities $\{p_1, \ldots, p_L\}$.

Set $x_k = \mathbf{w}^{\top} \mathbf{Z}_\ell$.

}
Return $\{x_k\}_{k=1}^K$.
\end{algorithm}

\begin{algorithm}[!htb]
\caption{Pre-compute ZANIM-LN-BART predictions at all proposed climate covariate values.}
\label{alg:ensemble_predictions}
\SetAlgoLined
\SetAlgoItemize
\SetKwInput{KwInput}{Input}
\SetKwInput{KwOutput}{Output}
\KwInput{
Observed modern climate $\mathbf{x}^m$, number of proposals $K$, and the posterior draws of the taxon-specific regression tree parameters, $\Psi^{(\mathrm{c}), r}_j=\{(\mathcal{T}^{(\mathrm{c}, r)}_{hj}, \Lambda^{(r)}_{hj})\}_{h=1}^{M_{\mathrm{c}}}$ and $\Psi^{(0, r)}_j=\{(\mathcal{T}^{(0, r)}_{hj}, \mathcal{M}^{(r)}_{hj})\}_{h=1}^{M_0}$, for $j \in \{1,\ldots, d\}$ taxa and $r\in\{1, \ldots, R\}$ posterior samples.
}
\KwOutput{
Predictions of the taxon-specific regression tree ensembles for the compositional probabilities $\{\theta^{(r)}_{kj}\}$ and structural zero probabilities $\{\zeta^{(r)}_{kj}\}$ across the $R$ posterior draws at the proposal values $\{\mathbf{x}_{k}\}_{k=1}^K$.}

Generate $K$ uniform draws $\{\mathbf{x}_{k}\}_{k=1}^K$ on the convex hull of $\mathbf{x}^m$ using Algorithm~\ref{alg:runif_convexhull}.

\For{$r$ from $1$ to $R$} {

\For{$k$ from $1$ to $K$} {

Compute the taxon-specific compositional probability parameters of the regression tree ensembles via
\[
\theta^{(r)}_{kj} \coloneqq \frac{f^{(\mathrm{c})}_j\left(\mathbf{x}_k; \Psi^{(\mathrm{c}, r)}_j\right)}{\sum_{\ell=1}^d  f^{(\mathrm{c})}_\ell\left(\mathbf{x}_k; \Psi^{(\mathrm{c}, r)}_\ell\right) },\]
where
\[\log\left\lbrack f_j^{(\mathrm{c})}\left(\mathbf{x}_k; \Psi^{(\mathrm{c}, r)}_j\right)\right\rbrack =
\sum_{h=1}^{M_{\mathrm{c}}}g\left(\mathbf{x}_k; \mathcal{T}^{(\mathrm{c}, r)}_{hj}, \Lambda^{(r)}_{hj}\right), \quad j\in\{1\ldots, d\}.
\]

Compute the taxon-specific structural zero probability parameters of the regression tree ensembles via
\[
\zeta^{(r)}_{kj} \coloneqq \Phi\left\lbrack f^{(0)}_j\left(\mathbf{x}_k; \Psi^{(0, r)}_j\right)\right\rbrack,\]
where
\[f_j^{(0)}\left(\mathbf{x}_k; \Psi^{(0, r)}_j\right) = \sum_{h=1}^{M_{0}}g\left(\mathbf{x}_k; \mathcal{T}^{(0, r)}_{hj}, \mathcal{M}^{(r)}_{hj}\right), \quad j\in\{1\ldots, d\}.
\]
}
}

Return the predictions $\{\theta^{(r)}_{kj}\}$ and $\{\zeta^{(r)}_{kj}\}$ of the regression tree ensembles and the associated proposal values $\{\mathbf{x}_k\}_{k=1}^K$.
\end{algorithm}

As discussed in Section~\ref{sec:sir}, our implementation of the SIR algorithm begins with a pre-processing step that can be decomposed into two sequential tasks.
First, $K$ proposal values $\{\mathbf{x}_k\}_{k=1}^K$ are generated uniformly over the convex hull of the modern climate observations $\mathbf{x}^m$.
To that end, we note that Algorithm~\ref{alg:runif_convexhull} describes our method to generate uniform random points inside the convex hull of a given generic matrix $\mathbf{X}$, for notational simplicity, by sampling from the components of the corresponding Delaunay triangulation in proportion to their volumes.
Next, the taxon-specific regression tree ensembles of the compositional probabilities and structural zero probabilities are evaluated at each proposal value, and stored across all posterior draws under the ZANIM-LN-BART model.
Algorithm~\ref{alg:ensemble_predictions} summaries these steps.
The output of Algorithm~\ref{alg:ensemble_predictions} is subsequently used for the SIR approach in Algorithm~\ref{alg:sir_multiple_imputation} during the importance weighting and resampling steps for each fossil sample.
For simplicity, Algorithm~\ref{alg:sir_multiple_imputation} is written in terms of a single fossil pollen sample $\mathbf{y}_s$.

\begin{algorithm}[!htb]
\caption{SIR with multiple imputation to approximate the inverse posterior in \eqref{eq:marginal_cut_posterior}.}
\label{alg:sir_multiple_imputation}
\SetAlgoLined
\SetAlgoItemize
\SetKwInput{KwInput}{Input}
\SetKwInput{KwOutput}{Output}
\KwInput{
Fossil pollen counts $\mathbf{y}_s$, posterior draws of the covariance matrix $\{\Sigma^{(r)}_U\}_{r=1}^R$, and quantities pre-computed via Algorithm~\ref{alg:ensemble_predictions}: predictions for the compositional probabilities, $\{\theta^{(r)}_{kj}\}_{r=1}^R$, and structural zero probabilities, $\{\zeta^{(r)}_{kj}\}_{r=1}^R$, along with associated climate proposal values $\{\mathbf{x}_{k}\}_{k=1}^K$.
}
\KwOutput{$R$ posterior draws of the unobserved climate $\mathbf{x}^{(r)}_s$ associated to the fossil pollen sample $\mathbf{y}_s$.}
\For{$r$ from $1$ to $R$} {
\For{$k$ from $1$ to $K$} {
Compute the importance weights $w_k = \hat{p}(\mathbf{y}_s \mid \bm{\theta}^{(r)}_{k}, \bm{\zeta}^{(r)}_{k}, \bm{\Sigma}^{(r)}_U)$ by obtaining an unbiased estimate of the ZANIM-LN-BART likelihood using Algorithm~\ref{alg:unbiased_estimate_zanim_ln} with $T=1$.
}
Resample a single value $\mathbf{x}_s^{(r)}$ from $\{\mathbf{x}_k\}_{k=1}^K$ with probabilities proportional to $\{w_k\}_{k=1}^K$.
}
Return $\{\mathbf{x}_s^{(r)}\}_{r=1}^{R}$.
\end{algorithm}

Finally, Algorithm~\ref{alg:ess} presents the alternative cESS approach within the na{\"{i}}ve cut algorithm to approximate the inverse posterior distribution of the past climate, $\mathbf{x}_s$, given the corresponding compositional pollen counts, $\mathbf{y}_s$.
Like our SIR approach, our cESS approach restricts the underlying prior to the domain of the modern climate covariates.
The ESS algorithm follows a similar structure, without requiring the evaluation of the terms in lines~\ref{step:loglike_thresh} and \ref{step:accept} associated with the logistic approximation of the indicator function \citep{Maatouk2026} and with $\mathbf{x}_s^{(1)}$ initialised from a non-truncated multivariate normal prior.
Recall that the mean and covariance parameters of the (truncated) multivariate normal prior under the (c)ESS approaches are set at their empirical values for the modern climate variables.

\begin{algorithm}[!htb]
\caption{cESS within the na{\"{i}}ve cut algorithm to approximate the inverse posterior in \eqref{eq:marginal_cut_posterior}.}
\label{alg:ess}
\SetAlgoLined
\SetAlgoItemize
\SetKwInput{KwInput}{Input}
\SetKwInput{KwOutput}{Output}
\SetKwInput{KwInit}{Initialise}
\SetKw{KwGoTo}{Go to}
\KwInput{Fossil pollen counts $\mathbf{y}_s$, draws of $\{\bm{\Psi}^{(r)}\}_{r=1}^R$, half-space representation of the convex-hull $\mathcal{C}$, i.e., the matrix of coefficients $\mathbf{A} \in \mathbb{R}^{q\times p}$ with rows $\mathbf{a}_k\in\mathbb{R}^{p\times 1}$ and the vector of offsets $\mathbf{b} \in \mathbb{R}^{q\times 1}$, the hyperparameter $\eta$ that controls the smooth approximation of the indicator function, and the empirical mean vector $\bm{\mu}$ and covariance matrix $\bm{\Sigma}$ of the modern climate variables.}
\KwInit{$\mathbf{x}_s^{(1)}$ by sampling from its prior $\operatorname{TMV}_d\lbrack\mathcal{C},\bm{\mu}, \bm{\Sigma}\rbrack$.}
\For{$r$ from $1$ to $R$} {

    Compute the predictions of the regression tree ensembles at a given climate covariate vector $\mathbf{x}_s^{(r)}$:
    \[\theta^{(r)}_{sj} \coloneqq \frac{f^{(\mathrm{c})}_j\left(\mathbf{x}^{(r)}_s; \Psi^{(\mathrm{c}, r)}_j\right)}{
    \sum_{\ell=1}^d f^{(\mathrm{c})}_\ell\left(\mathbf{x}^{(r)}_s; \Psi^{(\mathrm{c}, r)}_\ell\right)},\qquad
    \zeta^{(r)}_{sj} \coloneqq \Phi\left\lbrack f^{(0)}_j(\mathbf{x}^{(r)}_s; \Psi^{(0, r)}_j)\right\rbrack,\]
    where
    \[\log\left\lbrack f_j^{(\mathrm{c})}\left(\mathbf{x}^{(r)}_s; \Psi^{(\mathrm{c}, r)}_j\right)\right\rbrack =
    \sum_{h=1}^{M_{\mathrm{c}}}g\left(\mathbf{x}^{(r)}_s; \mathcal{T}^{(\mathrm{c}, r)}_{hj}, \Lambda^{(r)}_{hj}\right), \quad j\in\{1\ldots, d\},\]
    and
    \[f_j^{(0)}\left(\mathbf{x}^{(r)}_s; \Psi^{(0, r)}_j\right) = \sum_{h=1}^{M_{0}}g\left(\mathbf{x}^{(r)}_s; \mathcal{T}^{(0, r)}_{hj}, \mathcal{M}^{(r)}_{hj}\right), \quad j\in\{1\ldots, d\}.\]

    Draw an ellipse: $\bm{\nu} \sim \operatorname{Normal}\lbrack\bm{\mu}, \bm{\Sigma}\rbrack$.

    Compute the log-likelihood threshold:\label{step:loglike_thresh}
\[\log(\xi) = \log \left\lbrack
    \hat{p}\left(\mathbf{y}_s \mid \bm{\theta}^{(r)}_{s}, \bm{\zeta}^{(r)}_{s}, \bm{\Sigma}^{(r)}_U\right) \right\rbrack+ \sum_{k=1}^q \log\left( 1 + e^{-\eta (\mathbf{a}^\top_k \mathbf{x}_s^{(r)} - b_k)}\right)^{-1}  + \log(u),\]
    where $u \sim \operatorname{Uniform}\lbrack 0, 1\rbrack$ and the log-likelihood is estimated
using Algorithm~\ref{alg:unbiased_estimate_zanim_ln} with $T=100$.

    Draw the proposal:\label{step:ESS_proposal}
    \[\alpha \sim \operatorname{Uniform}\lbrack 0, 2\pi \rbrack, \quad \mathbf{x}_s^{\prime} = (\mathbf{x}_s^{(r)} - \bm{\mu}) \cos (\alpha) + (\bm{\nu} - \bm{\mu}) \sin (\alpha) + \bm{\mu}.\]

    Define the bracket: $\alpha_{\min} = \alpha - 2\pi$ and $\alpha_{\max} = \alpha$.

    Compute the predictions of the regression tree ensembles using the proposed climate covariates $\mathbf{x}_s^{\prime}$:
    \[\theta^{\prime}_{sj} \coloneqq \frac{f^{(\mathrm{c})}_j\left(\mathbf{x}^{\prime}_s; \Psi^{(\mathrm{c}, r)}_j\right)}{
    \sum_{\ell=1}^d f^{(\mathrm{c})}_\ell\left(\mathbf{x}^{\prime}_s; \Psi^{(\mathrm{c}, r)}_\ell\right)}, \qquad
    \zeta^{\prime}_{sj} \coloneqq \Phi\left\lbrack f^{(0)}_j\left(\mathbf{x}^{\prime}_s; \Psi^{(0, r)}_j\right)\right\rbrack,\]
    where
    \[\log\left\lbrack f_j^{(\mathrm{c})}\left(\mathbf{x}^{\prime}_s; \Psi^{(\mathrm{c}, r)}_j\right)\right\rbrack =
    \sum_{h=1}^{M_{\mathrm{c}}}g\left(\mathbf{x}^{\prime}_s; \mathcal{T}^{(\mathrm{c}, r)}_{hj}, \Lambda^{(r)}_{hj}\right), \quad j\in\{1\ldots, d\},\]
    and
    \[f_j^{(0)}\left(\mathbf{x}^{\prime}_s; \Psi^{(0, r)}_j\right) = \sum_{h=1}^{M_{0}}g\left(\mathbf{x}^{\prime}_s; \mathcal{T}^{(0, r)}_{hj}, \mathcal{M}^{(r)}_{hj}\right), \quad j\in\{1\ldots, d\}.\]

    \eIf{$\log \left\lbrack
    \hat{p}\left(\mathbf{y}_s \mid \bm{\theta}^{\prime}_{s}, \bm{\zeta}^{\prime}_{s}, \bm{\Sigma}^{(r)}_U\right) \right\rbrack
     +\sum_{k=1}^q \log\left( 1 + e^{-\eta (\mathbf{a}^\top_k \mathbf{x}_s^{\prime} - b_k)}\right)^{-1} > \log(\xi)$\label{step:accept}} {
     Accept: $\mathbf{x}_s^{(r+1)}=\mathbf{x}_s^{\prime}$.
    } {
        If $\alpha < 0$, then  $\alpha_{\min} = \alpha$ and $\alpha_{\max}=2\pi$. Otherwise, $\alpha_{\min}=0$ and $\alpha_{\max} = \alpha$.

        Draw a new angle: $\alpha \sim \operatorname{Uniform}\lbrack \alpha_{\min}, \alpha_{\max} \rbrack$.

        \KwGoTo line~\ref{step:ESS_proposal}.
    }
}
Return $\{\mathbf{x}_s^{(r)}\}_{r=1}^{R}$.
\end{algorithm}
 \section{Simulations for one-dimensional climate}\label{supp:add_sim}

In Section~\ref{sec:sim_studies}, we conducted simulations to assess the performance of our framework in settings with $p=2$ and $p=3$ climate variables.
In the absence of publicly available software for Bayesian palaeoclimate reconstruction when $p>1$, we restricted our attention to a comparison between the different sampling schemes (SIR, ESS, and cESS) for the reconstruction module.
Here, we present additional simulations for $p=1$ comparing the proposed methodology against two existing Bayesian palaeoclimate reconstruction methods which do not account for structural zeros: namely, the DM-GP and DM-BUMMER models introduced by \citet{Tipton2019} and \citet{Vasko2000}, respectively.
These methods follow the same modular inference approach as in our framework.
However, they exclusively use the ESS algorithm for the reconstruction module.
We also aim to to illustrate that ignoring structural zeros, as these methods do, can deteriorate the performance in terms of reconstructing past climates.

We assess the performance of the methods using similar metrics as discussed in Section~\ref{sec:sim_studies}.
However, for the scoring rule, we report the continuous rank probability score \citep[CRPS;][]{Gneiting2007}, which is the special case of the energy score (ES) when $p=1$.
In particular, given the posterior draws $\{x_s^{(r)}\}_{r=1}^{R}$ of the one-dimensional climate variable, the CRPS for the $s$-th observation is given by:
$\operatorname{CRPS}(x_s)=n_f^{-1}\sum_{r=1}^R \lvert x^{(r)}_s - x_s \rvert - 1/(2n_f^2)\sum_{r=1}^R\sum_{j=1}^R \lvert x^{(r)}_s - x^{(j)}_s\rvert$.

In these simulations, we consider $d=4$ taxa, $n_m=1{,}000$ samples for the modern data, and a single climate covariate $x^m_i$, since the current implementation available in the \texttt{BayesComposition} package is limited to $p=1$.
We simulate the covariate to resemble the MTCO variable of our application by generating $x^m_i \sim \operatorname{Normal}\left\lbrack \mu, \tau\right\rbrack$, with $\mu$ and $\tau$ being the mean and variance of the observed modern MTCO values.
To avoid simulating the compositional counts from the underlying distributional assumptions of our ZANIM-LN-BART model,
we simulate the compositional counts following the zero-and-$N$-inflated Dirichlet-multinomial distribution \citep{Menezes2026}.
Specifically, we generate $\mathbf{y}^m_i\sim \operatorname{Multinomial}\lbrack N_i, \bm{\vartheta}_i\rbrack$, where
the number of trials is generated as $N_i \sim \operatorname{Poisson}\lbrack 1000\rbrack$, and the multinomial probabilities are
$\vartheta_{ij}=z_{ij}\lambda_{ij}/\sum_{k=1}^dz_{ik}\lambda_{ik}$, with $z_{ij}\sim \operatorname{Bernoulli}\lbrack 1- \zeta_{ij}\rbrack$ and $\lambda_{ij}\sim\operatorname{Gamma}\lbrack \alpha_{ij}, 1\rbrack$, for observation $i\in\{1,\ldots, n_m\}$ and taxa $j\in\{1,\ldots, d\}$, where $\alpha_{ij}$ and $\zeta_{ij}$ are the count and structural zero probability parameters, respectively.
Following the assumption often made in pollen-based climate reconstruction that each pollen taxon has an optimum temperature \citep[see, e.g.,][]{Vasko2000}, we specify a unimodal Gaussian functional form on the compositional probabilities via $\alpha_{ij} = a_j \exp(-(x_i - b_j) /c_j^2 )$.
We fix the taxon-specific response parameters to represent taxa with distinct ecological preferences, using $\mathbf{a}=(4.3, 1.1,3.2,4.8)$, $\mathbf{b}=(-6,-5,-2,3)$, and $\mathbf{c}=(20,15,12,10)$, where $a_j$ is a scaling factor, $b_j$ is the optimum temperature of taxon $j$, and $c_j$ is a tolerance parameter.
Following \citet{SalterTownshend2012}, we define the structural zero probabilities through a power link function as $\zeta_{ij} = 1 - (\alpha_{ij}/\sum_{k=1}^d\alpha_{ik})^{\rho_j}$, with $\bm{\rho}=(0.3, 0.1, 0.4, 0.2)$.
This reparameterisation implies that structural zeros are more likely when pollen abundance is low, reflecting conditions where taxa are poorly supported in suboptimal temperature regions.
The $n_f=100$ samples from the fossil data set $\{\mathbf{y}_s, x_s\}$ were generated using the same recipe.
Although $x_s$ is unobserved in practical applications, we generate it in order to compute the predictive measures described in Section~\ref{sec:sim_studies}.

Table~\ref{tab:results_scenario1} presents the predictive performance of the methods.
Our framework based on the ZANIM-LN-BART model, regardless of the sampling scheme employed under the reconstruction module, consistently outperforms the DM-BUMMER (ESS) and DM-GP (ESS) methods across all predictive metrics.
This demonstrates that explicitly accounting for structural zeros, as we do in our framework, substantially improves reconstruction performance.
Among the three sampling strategies employed in the reconstruction module of our framework using the ZANIM-LN-BART model (SIR, ESS, and cESS), the cESS approach achieves the best overall performance.
Using a truncated normal prior for the missing covariates under cESS provides an improvement over the normal prior used with ESS, reducing the CRPS from $2.5926$ to $2.4526$, while maintaining coverage close to the nominal level for both sampling schemes.
Although SIR also achieves coverage close to the nominal level, its higher CRPS ($3.0616$) indicates a loss of predictive accuracy compared to both ESS and cESS.

\begin{table}[!htb]
\centering
\caption{
Comparison of the predictive performance of different pollen-based Bayesian reconstruction methods under the simulated scenario with $p=1$.
For each of the six replicates, predictive metrics were computed from the true $n_f=100$ past climate samples and then averaged across replicates.
The MSEP (MAE) assesses the accuracy of the posterior-predictive mean (median), whereas the CRPS is a proper scoring rule that evaluates the full posterior-predictive distribution. Smaller values of CRPS, MSEP, and MAE indicate better predictive performance, while coverage values close to the nominal level ($0.95$) indicate well-calibrated posterior uncertainty.
Values in bold indicate the best performing method under each predictive metric.}
\label{tab:results_scenario1}
\begin{tabular}{llrrrrr}
  \hline
Method & CRPS & MSEP & MAE & $95\%$ coverage \\
  \hline
ZANIM-LN-BART (SIR) & 3.062 & 67.601 & 4.227 & 0.969 \\
ZANIM-LN-BART (ESS) & 2.593 & 54.498 & 3.104 & 0.968 \\
ZANIM-LN-BART (cESS) & \textbf{2.453} & \textbf{47.961} & \textbf{2.986} &\textbf{0.967} \\
DM-GP (ESS) & 5.521 & 104.087 & 7.622 & 0.908 \\
DM-BUMMER (ESS) & 5.684 & 110.086 & 7.926 & 0.930 \\
   \hline
\end{tabular}
\end{table}

Panel \textbf{A} of Figure~\ref{fig:forward_reconstruction_scenario_1} shows, for a given, randomly chosen replicate, the posterior estimates of the true compositional probabilities under the models used to infer the pollen-climate relationships in the forward module.
Four representative examples of inverse posterior distributions arising from the reconstruction module are also shown in Panel \textbf{B}.
According to Panel \textbf{A}, the DM-BUMMER and DM-GP models clearly provide biased estimates of the taxon-specific compositional probabilities, while our ZANIM-LN-BART model can effectively recover the true values.
This is expected as both competing methods cannot handle structural zeros, resulting in estimates that are over/underestimated, depending on the taxon.
This poor fit is propagated to the reconstruction stage of the DM-BUMMER and DM-GP methods as illustrated in Panel \textbf{B} of Figure~\ref{fig:forward_reconstruction_scenario_1}.
In general, the inverse posterior distributions of the past climate $x_s$ obtained under both DM-BUMMER and DM-GP exhibit a large spread and are far from the true value of $x_s$. For instance, in the case where the observed fossil counts $\mathbf{y}_s$ contain no zeros (Panel \textbf{B}, top left), the resulting inverse posterior under DM-BUMMER and DM-GP is highly dispersed, whereas our method yields a distribution that is concentrated around the true value. A similar pattern is observed when one taxon has a single zero count (top right).
These results are consistent with the overall performance of the methods reported in Table~\ref{tab:results_scenario1}.

\begin{figure}[!htb]
    \centering
    \includegraphics[width=1.0\linewidth]{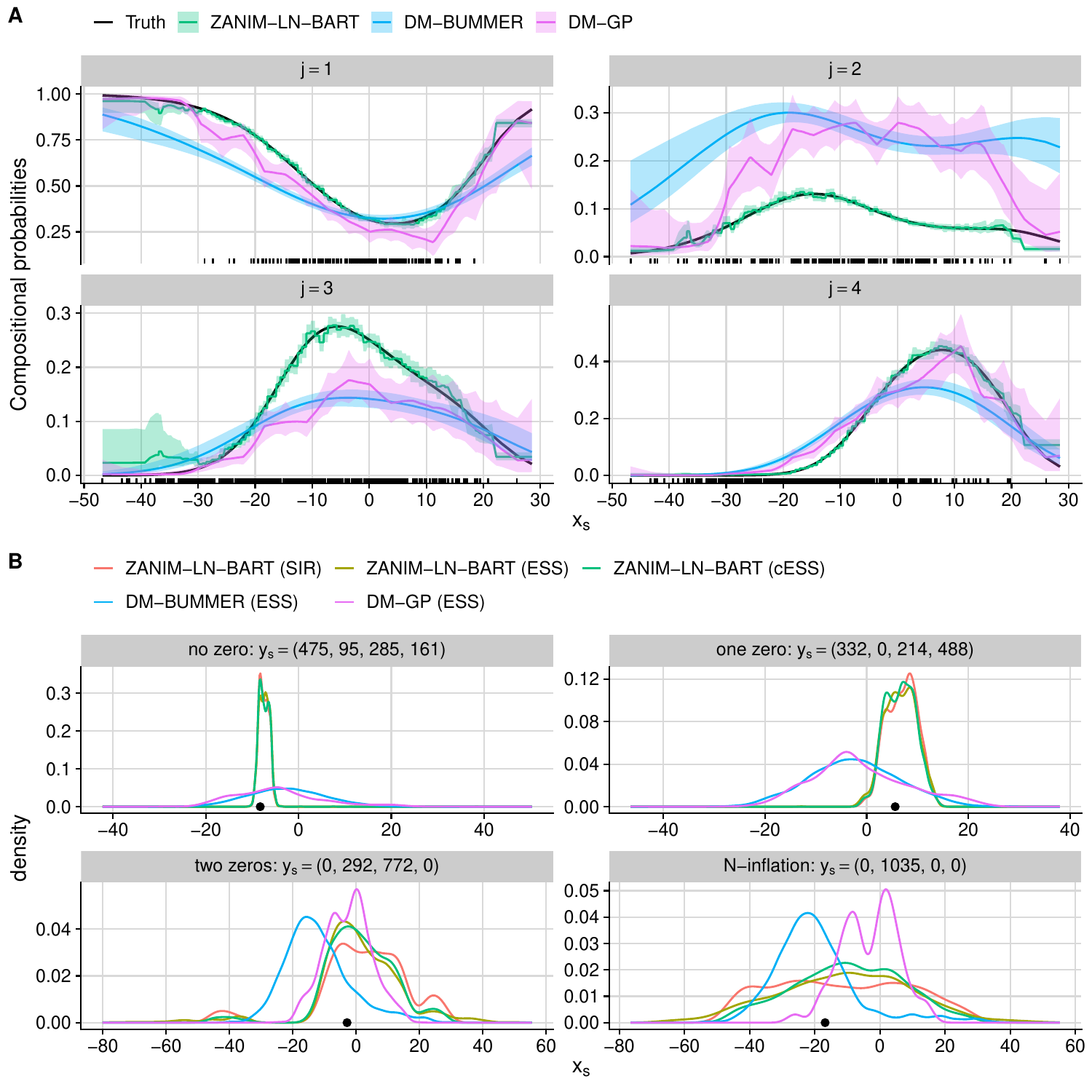}
    \caption{
    Results of the forward and reconstruction modules for a given replicate under the simulated scenario with $p=1$.
    Panel \textbf{A}: Posterior estimates of the true (black lines) pollen-taxa compositional probabilities $\alpha_{ij}/\sum_{k=1}^d\alpha_{ik}$ under the ZANIM-LN-BART, DM-BUMMER, and DM-GP models. The posterior median (lines) and $95\%$ credible intervals (shaded regions) are given in each case. The rugs along the $x$-axes represent samples where the observed modern pollen counts are zero.
    Panel \textbf{B}:
    Four examples of the reconstruction module showing kernel density estimates of the inverse posterior distributions under various methods. Each facet corresponds to a different fossil sample: no zeros, one zero, two zeros, and $N$-inflation. Our approach, based on the ZANIM-LN-BART model, employs three sampling schemes (SIR, ESS, and cESS) for the reconstruction module.
    The DM-BUMMER and DM-GP methods implemented in the \texttt{BayesComposition} package exclusively use the ESS scheme for the reconstruction stage.
    The black dots indicate the true values of $x_s$ in each case.}
    \label{fig:forward_reconstruction_scenario_1}
\end{figure}
 \section{Additional pollen-climate relationships heatmaps}\label{supp:extra_plots}

In Section~\ref{sec:app_forward_fit}, certain pollen-climate relationships uncovered by the ZANIM-LN-BART model fitted to the modern calibration data set were illustrated using heatmaps of the posterior median predictions of the $\theta_{ij}$, $\zeta_{ij}$, and $\vartheta_{ij}$ parameters.
These quantities correspond to the population-level compositional probabilities, population-level structural zero probabilities, and individual-level compositional probabilities, respectively.
These heatmaps were shown for just three representative pollen taxa, with just one pair among the $p=3$ available climate covariates examined in each case.
Specifically, Figure~\ref{fig:climate_predictions_chosen} showed GDD5 vs. AET/PET for the \textit{Pinus.D} taxon, MTCO vs. AET/PET for the \textit{Betula} taxon, and GDD5 vs. MTCO for the \textit{Juniperus} taxon.

In Figures~\ref{fig:climate_predictions_pinusd}, \ref{fig:climate_predictions_betula}, and \ref{fig:climate_predictions_juniperus}, we provide similar plots for the same three respective taxa, now showing all three covariate pairs in each case.
In effect, each set of plots partially reproduces a portion of Figure~\ref{fig:climate_predictions_chosen}.
However, showing the remaining covariate pairs which were not previously depicted allows us to consider the full three-dimensional grid over the modern climate space over which the predictions of the depicted quantities were generated, albeit over the three two-dimensional cross sections.
Broadly speaking, the conclusions that can be drawn from these plots are consistent with those of Figure~\ref{fig:climate_predictions_chosen}.
In particular, we note from Figure~\ref{fig:climate_predictions_pinusd} that \textit{Pinus.D} is generally abundant across the climate space, with the abundances peaking at GDD5 values between $1{,}000$ and $2{,}000$, and the structural zero probabilities remaining generally low.
We also note from Figure~\ref{fig:climate_predictions_betula} that \textit{Betula} does not thrive in warmer and moisture-limited environments, whereas Figure~\ref{fig:climate_predictions_juniperus} suggests that \textit{Juniperus} favours warmer climates.

However, the extra panels in these plots, relative to Figure~\ref{fig:climate_predictions_chosen}, reveal other interesting uncovered relationships.
For instance, Figure~\ref{fig:climate_predictions_pinusd} indicates that \textit{Pinus.D} has low abundance under extreme cold climates, as characterised by MTCO values below $-30^\circ\mathrm{C}$.
Moreover, the complete picture of the climate space predictions for \textit{Betula} in Figure~\ref{fig:climate_predictions_betula} indicate that the increased structural zero probabilities are primarily associated with AET/PET.
While the GDD5 vs. MTCO heatmap in Panel \textbf{A} shows higher structural zero probabilities at higher values of both variables, Panels \textbf{B} and \textbf{C}, which include AET/PET, reveal clear regions where this taxon is structurally absent in drier environments, particularly for AET/PET below 0.7, MTCO above $-10^\circ\mathrm{C}$, and GDD5 greater than $1{,}000$.
Finally, Figure~\ref{fig:climate_predictions_juniperus} provides additional evidence that \textit{Juniperus} strives in warmer climates, but augments this with the insight from Panel \textbf{B} and Panel \textbf{C} that, provided such warmth is available, it tolerates both wet and dry environments.
For completeness, we provide analogous plots for the $25$ remaining pollen taxa at \url{https://github.com/AndrMenezes/reproduce__palaeoclimate_zanim_ln_bart}.

\begin{figure}[!htb]
    \centering
    \includegraphics[width=1.0\linewidth]{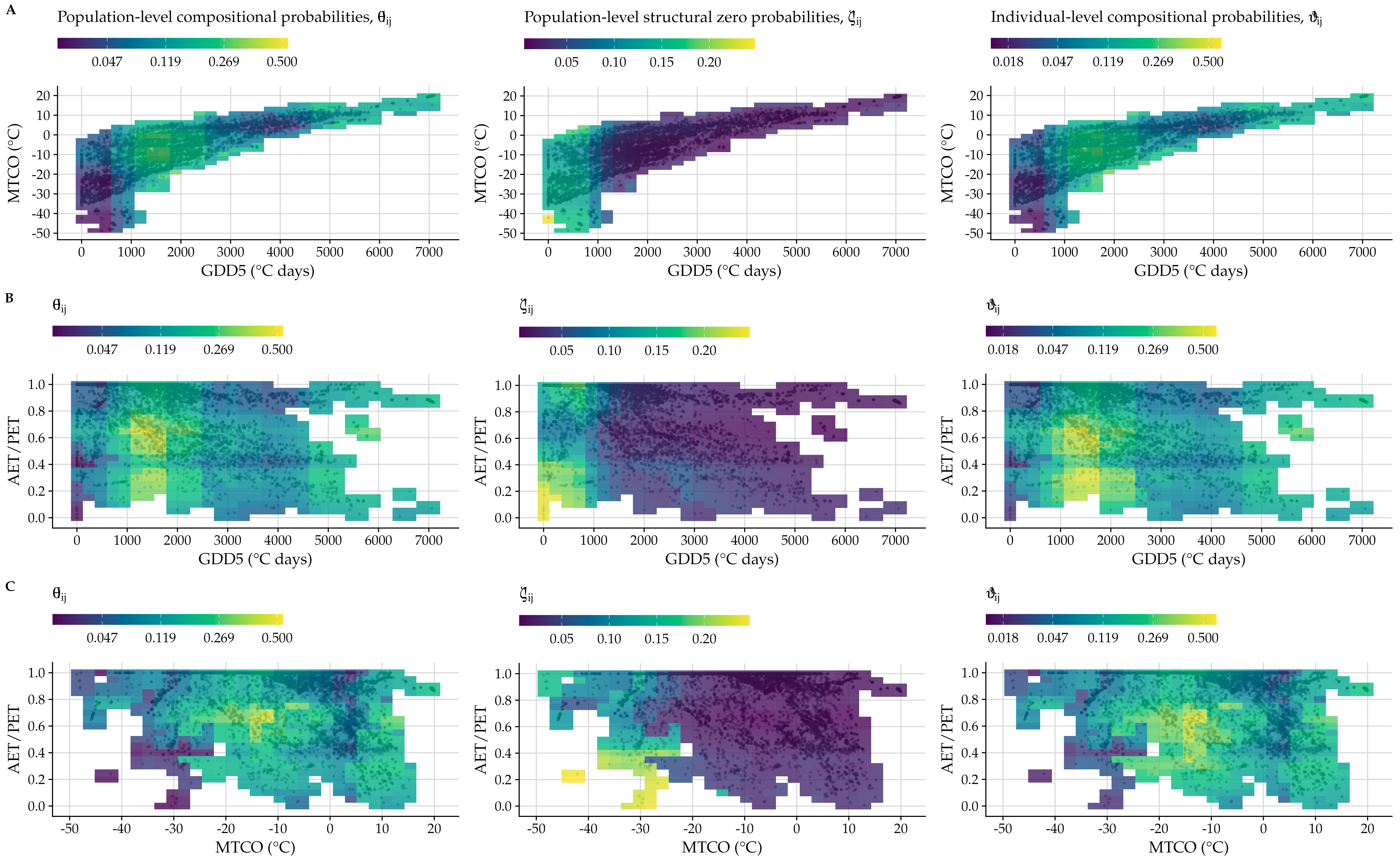}
    \caption{Heatmaps of the posterior median predictions of the ZANIM-LN-BART model parameters for the \textit{Pinus.D} taxon, with the population-level count probabilities ($\theta_{ij}$, left column), the population-level structural zero probabilities ($\zeta_{ij}$, middle column) and the individual-level count probabilities ($\vartheta_{ij}$, right column).
    The predictions were computed over a three-dimensional discrete grid of the climate covariates (GDD5, MTCO, and AET/PET) representing the modern climate sites at which pollen was actually observed.
    Black dots indicate the modern climate observations.
    Each panel represents a distinct pair of climate covariates, with each bivariate plot showing the model predictions for a given parameter.
    Darker colours correspond to lower values of the model parameters, with the black shading in the rightmost column, showing $\vartheta_{ij}$, corresponding to posterior medians of exactly zero.
    Panel \textbf{A}: GDD5 vs. MTCO. Panel \textbf{B}: GDD5 vs. AET/PET. Panel \textbf{C}: MTCO vs. AET/PET.}
    \label{fig:climate_predictions_pinusd}
\end{figure}

\begin{figure}[!htb]
    \centering
    \includegraphics[width=1.0\linewidth]{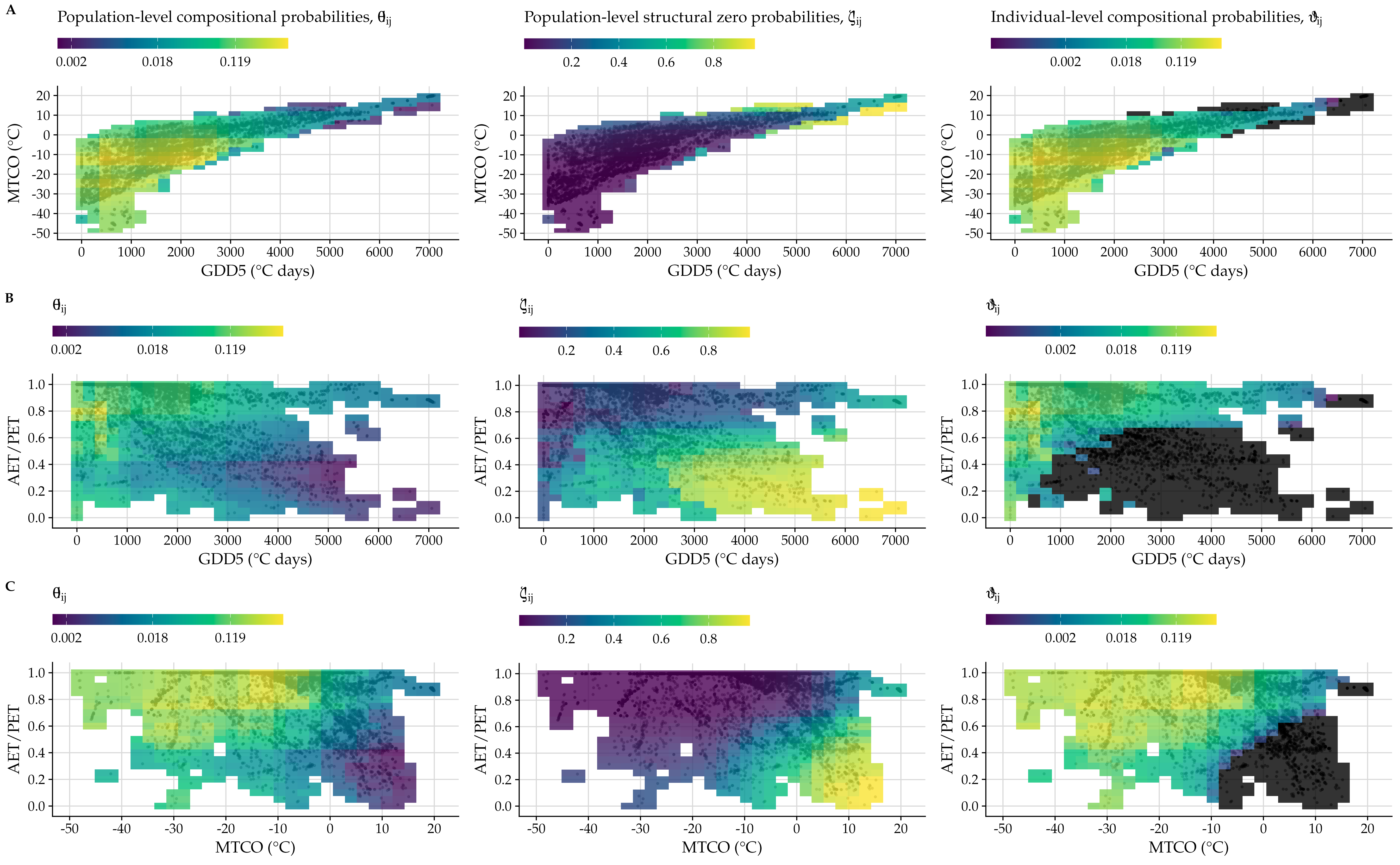}
    \caption{Heatmaps of the posterior median predictions of the ZANIM-LN-BART model parameters for the \textit{Betula} taxon, with the population-level count probabilities ($\theta_{ij}$, left column), the population-level structural zero probabilities ($\zeta_{ij}$, middle column) and the individual-level count probabilities ($\vartheta_{ij}$, right column).
    The predictions were computed over a three-dimensional discrete grid of the climate covariates (GDD5, MTCO, and AET/PET) representing the modern climate sites at which pollen was actually observed.
    Black dots indicate the modern climate observations.
    Each panel represents a distinct pair of climate covariates, with each bivariate plot showing the model predictions for a given parameter.
    Darker colours correspond to lower values of the model parameters, with the black shading in the rightmost column, showing $\vartheta_{ij}$, corresponding to posterior medians of exactly zero.
    Panel \textbf{A}: GDD5 vs. MTCO. Panel \textbf{B}: GDD5 vs. AET/PET. Panel \textbf{C}: MTCO vs. AET/PET.}
    \label{fig:climate_predictions_betula}
\end{figure}

\begin{figure}[!htb]
    \centering
    \includegraphics[width=1.0\linewidth]{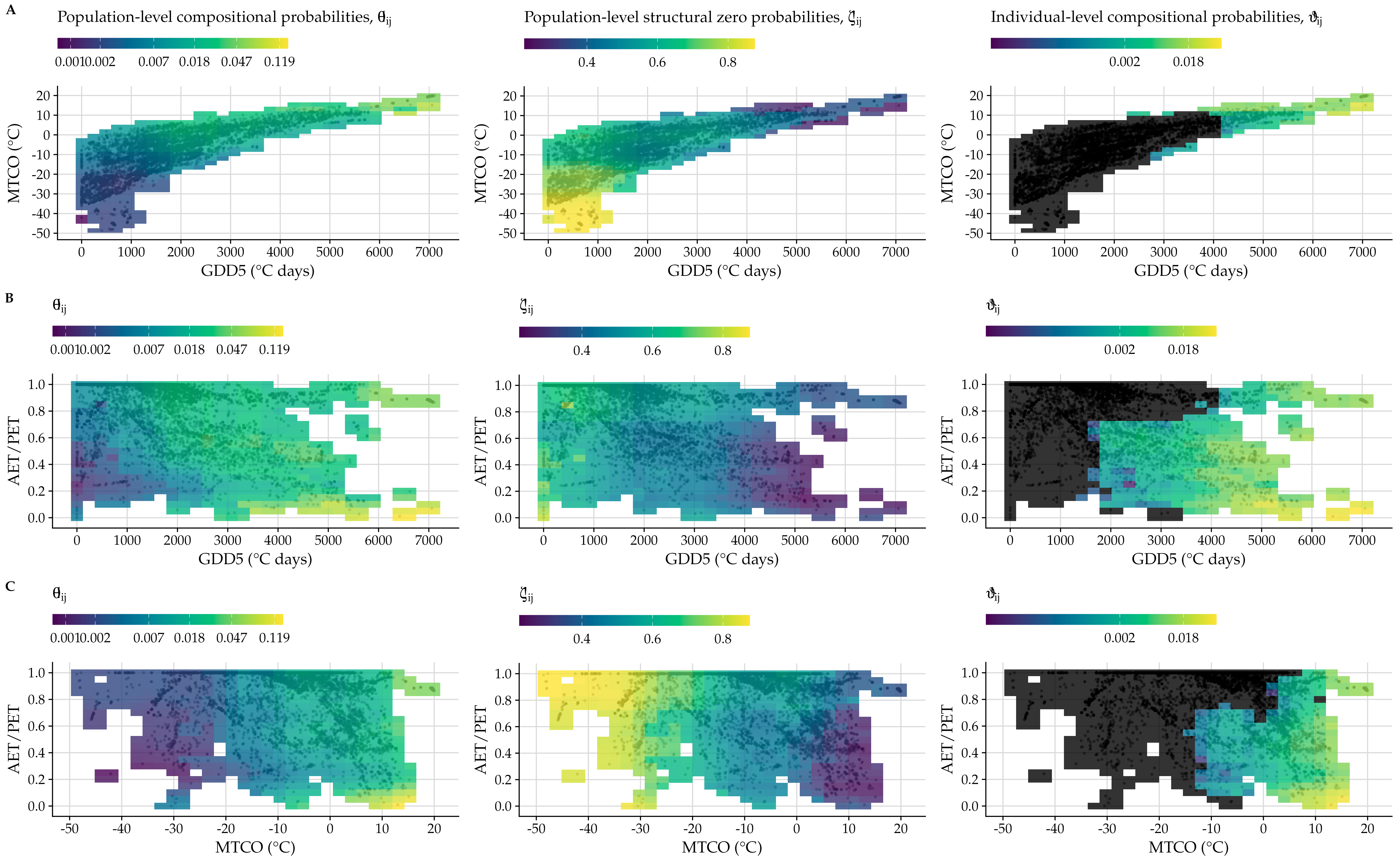}
    \caption{Heatmaps of the posterior median predictions of the ZANIM-LN-BART model parameters for the \textit{Juniperus} taxon, with the population-level count probabilities ($\theta_{ij}$, left column), the population-level structural zero probabilities ($\zeta_{ij}$, middle column) and the individual-level count probabilities ($\vartheta_{ij}$, right column).
    The predictions were computed over a three-dimensional discrete grid of the climate covariates (GDD5, MTCO, and AET/PET) representing the modern climate sites at which pollen was actually observed.
    Black dots indicate the modern climate observations.
    Each panel represents a distinct pair of climate covariates, with each bivariate plot showing the model predictions for a given parameter.
    Darker colours correspond to lower values of the model parameters, with the black shading in the rightmost column, showing $\vartheta_{ij}$, corresponding to posterior medians of exactly zero.
    Panel \textbf{A}: GDD5 vs. MTCO. Panel \textbf{B}: GDD5 vs. AET/PET. Panel \textbf{C}: MTCO vs. AET/PET.}
    \label{fig:climate_predictions_juniperus}
\end{figure} \section{\textsf{R} scripts}\label{supp:R_scripts}

Code with the implementations of our methodology is provided through the open source \textsf{R} package \texttt{zanicc} available from the GitHub repository at \url{https://github.com/AndrMenezes/zanicc}.
For efficient computation, the core parts of the MCMC algorithms are written in \textsf{C++} and functions are provided for both the forward and reconstruction modules.
In addition, we provide computational details of the pipelines to perform palaeoclimate reconstruction tasks in general --- using our modern data and the fossil data from Lago Grande di Monticchio, for illustrative purposes --- at \url{https://github.com/AndrMenezes/reproduce__palaeoclimate_zanim_ln_bart}.
Finally, we note that files containing the posterior draws of the ZANIM-LN-BART model fitted to the Northern Hemisphere modern calibration data can be made available, on request, by the first author of this paper. }

\end{document}